\pdfoutput=1
\documentclass[11pt,twoside,a4paper,cmspaper,final,collab]{cms-tdr}
\def\svnVersion{f870957}\def\svnDate{2024/02/19}\def\cmsCernNoTag{CERN-EP-2023-158}\def\cmsCernDate{\today}\def\cmsMessage{Published in the Journal of High Energy Physics as \href{http://dx.doi.org/10.1007/JHEP02(2024)106}{\doi{10.1007/JHEP02(2024)106}.}}
\begin{document}\cmsNoteHeader{HIN-21-010}

\newcommand{\dangle}[1]{\ensuremath{\langle\langle #1 \rangle\rangle}}
\newcommand{\sangle}[1]{\ensuremath{\langle #1 \rangle}}
\newcommand{\vc}[1]{\ensuremath{v_{2}\{#1\}}\xspace}
\newcommand{\ve}{\ensuremath{v_{2}}\xspace}
\newcommand{\vk}{\ensuremath{v_{2}\{2k\}}\xspace}
\newcommand{\vnk}{\ensuremath{v_{n}\{2k\}}\xspace}
\newcommand{\cnk}{\ensuremath{c_{n}\{2k\}}\xspace}
\newcommand{\cek}{\ensuremath{c_{2}\{2k\}}\xspace}
\newcommand{\vn}{\ensuremath{v_{n}}\xspace}
\newcommand{\vten}{\ensuremath{v_{n}\{10\}}\xspace}
\newcommand{\PbPb}{\ensuremath{\text{PbPb}}\xspace}
\newcommand{\probeold}{\ensuremath{\frac{v_{2}\{6\}-v_{2}\{8\}}{{v}_{2}\{4\}-v_{2}\{6\}}}\xspace}
\newcommand{\probenew}{\ensuremath{\frac{v_{2}\{8\}-v_{2}\{10\}}{{v}_{2}\{6\}-v_{2}\{8\}}}\xspace}
\cmsNoteHeader{HIN-21-010}
\title{Higher-order moments of the elliptic flow distribution in \texorpdfstring{\PbPb}{PbPb} collisions at \texorpdfstring{$\sqrtsNN = 5.02\TeV$}{sqrt(s[NN]) = 5.02 TeV}}
\date{\today}
\abstract{
The hydrodynamic flow-like behavior of charged hadrons in high-energy lead-lead collisions is studied through multiparticle correlations. The elliptic anisotropy values based on different orders of multiparticle cumulants, $v_{2}\{2k\}$, are measured up to the tenth order ($k = 5$) as functions of the collision centrality at a nucleon-nucleon center-of-mass energy of \sqrtsNN = 5.02\TeV. The data were recorded by the CMS experiment at the LHC and correspond to an integrated luminosity of 0.607\nbinv. A hierarchy is observed between the coefficients, with $v_{2}\{2\} > v_{2}\{4\} \gtrsim v_{2}\{6\} \gtrsim v_{2}\{8\} \gtrsim v_{2}\{10\}$. Based on these results, centrality-dependent moments for the fluctuation-driven event-by-event $v_{2}$ distribution are determined, including the skewness, kurtosis and, for the first time, superskewness. Assuming a hydrodynamic expansion of the produced medium, these moments directly probe the initial-state geometry in high-energy nucleus-nucleus collisions.
}
\hypersetup{
pdfauthor={CMS Collaboration},
pdftitle={Higher-order moments of the elliptic flow distribution in PbPb collisions at sqrt(s[NN]) = 5.02 TeV},
pdfsubject={CMS},
pdfkeywords={Quark gluon plasma, Hydrodynamics, Azimuthal anisotropies, Q-cumulants}}
\maketitle 
 
\section{Introduction}
In high-energy nucleus-nucleus collisions, a hot and dense state of strongly interacting quarks and gluons is created, the so-called quark-gluon plasma (QGP). Clear evidence for this state was obtained at the BNL RHIC~\cite{BRAHMS:2004adc,PHOBOS:2004zne,STAR:2005gfr,PHENIX:2004vcz}. Experiments at the CERN LHC, performed with nucleus-nucleus collisions at nucleon-nucleon center-of-mass energies up to more than twenty times greater than available for the RHIC studies, confirmed the main RHIC results with much larger event samples and kinematic ranges. Unexpectedly, a similar behavior has been observed in smaller systems formed in high-multiplicity proton-proton and proton-nucleus collisions~\cite{CMS:2016fnw,CMS:2013jlh}. One important feature of the QGP is its apparent collective hydrodynamic expansion. The initial spatial geometry of the overlap region of the colliding nuclei results in anisotropic pressure gradients that, in turn, are reflected in the azimuthal angle distribution of outgoing particles. This anisotropy has been used to perform detailed studies of the QGP formed at the LHC~\cite{ALICE:2010suc,ALICE:2011ab,ALICE:2014wao,ALICE:2016ccg,ATLAS:2011ah,ATLAS:2012at,ATLAS:2013xzf,CMS:2012xss,CMS:2012zex,CMS:2013wjq,CMS:2013bza,CMS:2015xmx,CMS:2017mzx}. Fluctuations in nuclear densities that appear due to the fluctuations in the position of nucleons within the incident nuclei also have a significant influence on the QGP expansion~\cite{PHOBOS:2006dbo,PHOBOS:2007vdf,Alver:2010gr,ALICE:2011ab,ATLAS:2012at}. The effect of these geometry fluctuations on the observed particle anisotropy can give insight into the early-stage dynamics of the collisions~\cite{Bhalerao:2011yg,Retinskaya:2013gca,Giacalone:2017uqx}.

\begin{figure}[ht!]
     \centering
  \includegraphics[width=0.65\textwidth]{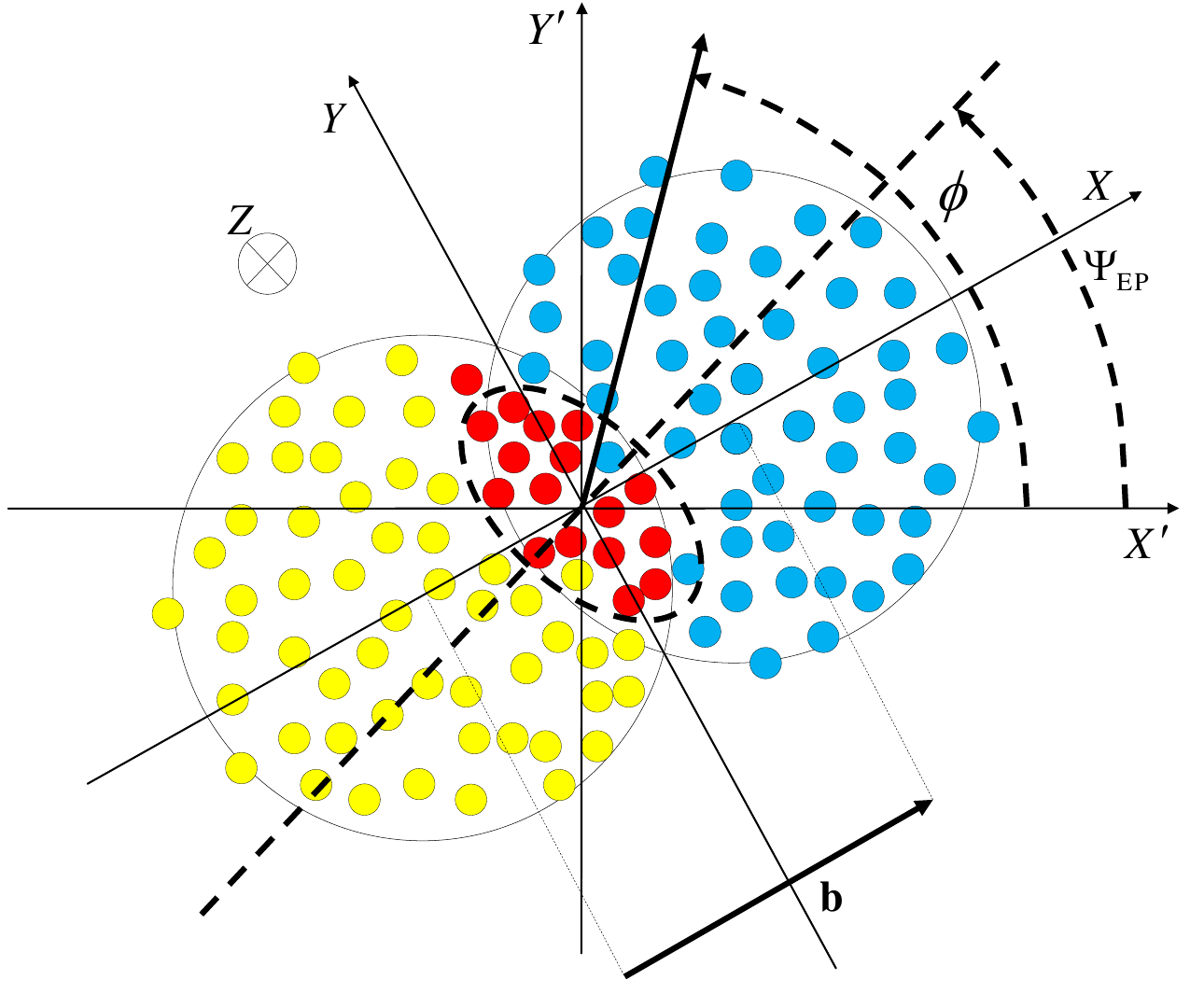}
  \caption{\label{fig:1} A schematic view of a non-central nucleus-nucleus collision in the transverse plane.}
\end{figure}

Figure~\ref{fig:1} schematically depicts a collision of two nuclei in the transverse $X'-Y'$ plane that is perpendicular to the beam axis $Z$, in the laboratory frame. The impact parameter vector $\mathbf{b}$, which is not experimentally accessible, connects the centers of the colliding nuclei in the transverse plane. The symmetry plane of the collision, denoted $X-Y$, defined in terms of the beam direction and $\mathbf{b}$, is randomly oriented with respect to the $X'-Y'$ plane. In this schematic figure, $\phi$ denotes the azimuthal angle of one of the outgoing particles. The anisotropic flow is quantified in terms of coefficients in the Fourier expansion of the $\phi$ dependence of the invariant yield of particles relative to the event-plane angle $\Psi_{\mathrm{EP}}$,
\begin{equation}
\label{FE}
 E\frac{\rd^{3}N}{\rd{p^{3}}} = \frac{1}{2\pi} \frac{\rd^{2}N}{\pt \rd\pt dy} \bigg[ 1 + \sum^{\infty}_{n=1}2v_{n} \cos[n(\phi-\Psi_{\mathrm{EP}})]\bigg],
\end{equation}
where $E$ is the energy of the particle, \pt its transverse momentum, and $y$ its rapidity. The event-plane angle $\Psi_{\mathrm{EP}}$ is defined to be in the direction of maximum outgoing particle density~\cite{Voloshin:1994mz} and is, on average, the same as the rotation angle of the symmetry plane with respect to the laboratory frame. The direction of $\Psi_{\mathrm{EP}}$ fluctuates about the symmetry plane rotation angle because of fluctuations in the initial state geometry and resolution effects arising from finite particle multiplicities. The elliptic flow harmonic \ve is the leading term of the Fourier series expansion of the azimuthal angle distribution in the event plane frame.

Many methods have been developed to study the anisotropic expansion of the system formed in nuclear collisions~\cite{Ollitrault:1993ba,Voloshin:1994mz,Poskanzer:1998yz}. The cumulant method was introduced in Refs.~\cite{Borghini:2000sa,Borghini:2001vi} based on multiparticle correlations which do not require an explicit measure of the event-plane angle. In this method, $v_{n}$ values can be extracted from \cnk cumulant coefficients, where $n$ is the order of the Fourier expansion term and $2k$ is the number of particles used to determine the correlation. For example, \vc{2}, \vc{4}, and \vc{6} are referred to as the 2-, 4-, and 6-particle cumulant based values for the elliptic flow harmonic, respectively. Nonflow, short-range correlations arising from jets and resonance decays, can be suppressed by correlating four or more particles. It is then possible to determine the central moments of the event-by-event, fluctuation-driven, \ve distribution based on the \vk values. The 3rd, 4th, and 5th central moments are called the skewness ($s$), kurtosis ($\kappa$), and superskewness ($p$) respectively. The skewness describes the overall asymmetry of the \ve distribution, while the superskewness gives a measure of the asymmetry of the tails. The kurtosis describes the peakedness of the center, or equivalently, the relative contribution (heaviness) of the distribution tails. In a hydrodynamic expansion the \ve is proportional to the initial-state eccentricity $\epsilon_{2}$, and therefore $\epsilon_{2}$ fluctuations produce non-Gaussian \ve distributions~\cite{Yan:2014nsa}. Thus, the central moments can be related to the initial state geometry and reveal its detailed structure.

In this paper, the cumulant based values are determined using the Q-cumulant method in which one calculates multiparticle cumulants in terms of moments of Q-vectors~\cite{Bilandzic:2010jr}. Determining the cumulant based \vk ($k=1,2,3\dots$) values of the \ve distribution is a way to study the fluctuation behavior related to the initial-state geometry~\cite{Borghini:2001vi}. For Gaussian fluctuations, the higher central moments vanish and the \vk values with $k > 1$ are all expected to have the same value~\cite{Voloshin:2007pc}. The measured higher-order \vk ($k = 2,3,4$) values, however, show a fine splitting~\cite{CMS:2017glf,ALICE:2018rtz}, $v_{2}\{4\} \gtrsim v_{2}\{6\} \gtrsim v_{2}\{8\}$, that can be attributed to a non-Gaussian behavior~\cite{Yan:2014nsa}. In Ref.~\cite{Giacalone:2016eyu}, it is noted that the main signature of non-Gaussian fluctuations is a nonzero skewness of the \ve distribution. This is used to suggest, as described in Ref.~\cite{Giacalone:2016eyu}, a hydrodynamic probe based on the ratio $(\vc{6}-\vc{8})/(\vc{4}-\vc{6})$. The basic premise of a hydrodynamic probe is that the observed azimuthal angle correlations that appear in the bulk medium can be related to the initial-state geometry.
 
We present for the first time a measurement of $v_{2}\{10\}$, derived from ten-particle cumulants, for lead-lead (\PbPb) collisions at a center-of-mass energy per nucleon pair of \sqrtsNN = 5.02\TeV. The data used in this analysis correspond to an integrated luminosity of 0.607\nbinv and were recorded with the CMS detector at the LHC. This enables the development of a new hydrodynamic probe using the \vc{10} value and the corresponding lower-order \vk values. Both the original and the new hydrodynamic probes should be independent of collision centrality if moments of the \ve distribution with order higher than the skewness are negligible~\cite{Giacalone:2016eyu}. Centrality refers to the percentage of the total inelastic hadronic nucleus-nucleus cross section~\cite{Miller:2007ri}, with 0\% corresponding to the maximum overlap of the colliding nuclei. This connection between the centrality dependence of the hydrodynamic probes and higher order moments of the \ve distribution is investigated by measuring both the centrality dependence itself and also by using the cumulants to extract the higher order moments directly.

This paper is organized as follows. In Section~\ref{method} and in Appendix~\ref{sec:appA}, we derive formulas for the \vc{10} harmonic and its statistical uncertainty. Section~\ref{newProbe} and Appendices~\ref{sec:appB}, and \ref{sec:appC} present derivations of the formulas used in this analysis to calculate the hydrodynamic probe ratios. In Section~\ref{moments}, we introduce and derive formulas for the standardized moments (denoted as $\gamma^{\text{exp}}_{i}$, $i=1,2,3$) and the corrected moments (denoted as $\gamma^{\text{exp}}_{i\text{, corr}}$, $i=1,2,3$) of the fluctuation-driven, event-by-event \ve distribution. Section~\ref{expData}, \ref{res}, and \ref{sum} present the experimental details, results, and summary, respectively. The numerical values of the results for this analysis are tabulated in the HEPDATA record~\cite{hepdata}.

\section{Analysis procedure}
\label{method}

A version of the cumulant method, the Q-cumulant method, was introduced in Ref.~\cite{Bilandzic:2010jr} and is based on the flow vector $Q_{n}$, with $Q_{n} = {\sum^{M}_{j=1}}\re^{in\phi_{j}}$. Here, $n$ is the flow harmonic order, $M$ denotes the event multiplicity, \ie, the number of analyzed tracks in the given event, and $\phi_{j}$ is the azimuthal angle of the $j$th track measured in the laboratory frame.

The $2m$-particle azimuthal angle correlators $\sangle{2m}$~\cite{Bilandzic:2010jr} are expressed through the flow vector $Q_{n}$ with
  \begin{equation}
  \label{single-event}
\sangle{2m} = \sangle{ \re^{in(\phi_{1}+\dots+\phi_{m}-\phi_{m+1}- \dots -\phi_{2m})} } = \frac{(M-2m)!}{M!}{\sum^{M}_{j_{1}\neq \dots \neq j_{2m}=1}}\re^{in(\phi_{j_{1}}+\dots+\phi_{j_{m}}-\phi_{j_{m+1}}-\dots-\phi_{j_{2m}})}.
  \end{equation}
Here the summation is over all unique track indices. The weighted mean of the $2m$-particle azimuthal angle correlator $\dangle{2m}$ is defined as
\begin{equation}
\label{2mCumul}
\dangle{2m} = \frac{\sum^{\text{events}}_{j}(W_{\sangle{2m}})_{j}\sangle{2m}_{j}}{\sum^{\text{events}}_{j}(W_{\sangle{2m}})_{j}}, \; m=1,\dots,k,
\end{equation}
where the double brackets $\dangle{\dots}$ denote a weighted average over all events within a given class~\cite{Bilandzic:2012wva}. The weight is taken as the number of distinct $2m$-particle combinations that can be formed for an event~\cite{Bilandzic:2012wva}:
\begin{equation}
\label{weight}
W_{\sangle{2m}} = \prod^{2m-1}_{j=0}(M-j) = \frac{M!}{(M-2m)!}.
\end{equation}
The Q-cumulant $c_{n}\{2k\}$ values are multivariate polynomial functions of the $\dangle{2m}$ values. The relationship between the \cnk values (up to the 8th order) and corresponding $\dangle{2m}$ 
correlations are known~\cite{Bilandzic:2010jr,Bilandzic:2012wva}. A recursion relationship for the \cnk values~\cite{Nadderd:2021fst} is given with
\begin{equation}
\label{Q-general}
c_{n}\{2k\} = \dangle{ 2k } - {\sum^{k-1}_{m=1}} \binom{k}{m} \binom{k-1}{m} \dangle{2m} c_{n}\{2k-2m\}.
\end{equation}
In Refs.~\cite{Bilandzic:2012wva,Wang:1991qh} it is shown that
\begin{equation}
  \label{2kVn}
\dangle{ 2k } = \sangle{ v_n^{2k} }.
\end{equation}
Based on Eqs.~(\ref{Q-general}) and (\ref{2kVn}), one finds
\begin{equation}
  \label{Vn}
v_n = \sqrt[2k]{a_{2k}^{-1}c_n\{2k\}},
\end{equation}
where the $a_{2k}$ coefficients can be determined using the recursion relationship
\begin{equation}
  \label{a2k}
a_{2k} = 1 - {\sum^{k-1}_{m=1}} \binom{k}{m} \binom{k-1}{m} a_{2k-2m}, \quad \text{with $a_{2} = 1$}.
\end{equation}
In the ideal case, all $2k$-particle cumulants \cnk with $k > 1$ result in the same flow harmonic value $v_n$. But, non-Gaussian flow fluctuations and non-flow correlations result in a splitting of the \cnk values with $k > 1$. In this case, the \vnk values will depend on the cumulant orders as
\begin{equation}
  \label{Vn2k}
v_n\{2k\} \equiv \sqrt[2k]{a_{2k}^{-1}c_n\{2k\}}.
\end{equation}
Details concerning the Q-cumulant method up to the $k=4$ order can be found in Refs.~\cite{Bilandzic:2010jr,Bilandzic:2012wva}. In the case where $k=5$, the above equations give the relationship between the $v_{n}\{10\}$ magnitude and the corresponding $c_{n}\{10\}$ value, with
\begin{multline}
  \label{Cumul10}
      c_{n}\{10\} = \dangle{10} - 25 \dangle{2} \dangle{8} - 100 \dangle{4} \dangle{6} + 400 \dangle{6} \dangle{2}^{2} + 900 \dangle{2} \dangle{4}^{2} \\
      - 3600 \dangle{4} \dangle{2}^{3} + 2880 \dangle{2}^{5},
\end{multline}
and
\begin{equation}
  \label{FCoeff}
      v_{n}\{10\} = \sqrt[10]{\frac{1}{456}c_{n}\{10\}}.
\end{equation}
Formulas for the analytic calculations of the statistical uncertainties in the $v_n\{2k\}$ ($k=1,\dots,4$) magnitudes are given in Ref.~\cite{Nadderd:2021fst}. Appendix~\ref{sec:appA} discusses the analytical calculation of the statistical uncertainty for the \vc{10} value.

\section{Hydrodynamic probes}
\label{newProbe}

A compact formulation for the cumulant expansion is given through the formalism of generating functions. The Fourier--Laplace transform of the symmetry plane elliptic harmonic vector $\mathbf{v}_{2}=v_{x}\mathbf{e}_{x}+v_{y}\mathbf{e}_{y}$ is $\sangle{ \re^{\mathbf{l}\boldsymbol{\cdot} \mathbf{v_{2}}} }$, with a vector variable $\mathbf{l} = l_{x}\mathbf{e}_{x}+l_{y}\mathbf{e}_{y}$. Here, $\mathbf{e}_{x}$ is the unit vector in the direction of the impact parameter, while $\mathbf{e}_{y}$ is the unit vector perpendicular to $\mathbf{b}$. The angle brackets $\sangle{}$ indicate an event average that, for this analysis, involves an average over events within a centrality class. As the distribution of ($v_{x},v_{y}$) is symmetric under $v_{y}\to  - v_{y}$, the mean value of the $x-$ and $y-$component of $\mathbf{v_{2}}$ are $\sangle{ v_{x} } \equiv \overline{v}_{2}$, and  $\sangle{ v_{y} } = 0$. The variances are $\sigma^{2}_{x} = \sangle{ (v_{x}-\sangle{ v_{x} })^{2}}$ and $\sigma^{2}_{y} = \sangle{ v^{2}_{y} }$. Higher-order central moments up to the kurtosis are defined in Ref.~\cite{Bhalerao:2018anl} in terms of the components of the elliptic harmonic vector. The expressions found in Ref.~\cite{Bhalerao:2018anl} are reproduced here, expanded to include terms related to the superskewness:
\begin{equation}
  \label{momDef}
  \begin{aligned}
    s_{30} &= \langle (v_{x}-\sangle{ v_{x} })^{3} \rangle, \\
    s_{12} &= \sangle{ (v_{x}-\sangle{ v_{x} })v^{2}_{y} }, \\
    \kappa_{40} &= \sangle{ (v_{x}-\sangle{ v_{x} })^{4} } - 3\sigma^{4}_{x}, \\
    \kappa_{22} &= \sangle{ (v_{x}-\sangle{ v_{x} })^{2}v^{2}_{y} } - \sigma^{2}_{x}\sigma^{2}_{y}, \\
    \kappa_{04} &= \sangle{ v^{4}_{y} } - 3\sigma^{4}_{y}, \\
    p_{50} &= \langle (v_{x}-\sangle{ v_{x} })^{5} \rangle - 10\sigma^{2}_{x}s_{30}, \\
    p_{32} &= \langle (v_{x}-\sangle{ v_{x} })^{3}v^{2}_{y} \rangle - \sigma^{2}_{y}s_{30} - 3\sigma^{2}_{x}s_{12}, \quad \text{and} \\
    p_{14} &= \langle (v_{x}-\sangle{ v_{x} })v^{4}_{y} \rangle - 6\sigma^{2}_{y}s_{12}.
  \end{aligned}
\end{equation}
In Ref.~\cite{Giacalone:2016eyu}, the generating function of the cumulant is defined as $\ln\sangle{ \re^{\mathbf{l}\cdot \mathbf{v_{2}}} }$. If one expands this generating function up to the 4th power in $(l_{x},l_{y})$ one gets the expression in terms of central moments of the $(v_{x},v_{y})$ distribution, that includes the $\sigma$, $s$, and $\kappa$ terms~\cite{Bhalerao:2018anl}. By extending this expansion up to the 5th power in $(l_{x}, l_{y})$, we obtain
\begin{multline}
  \label{expGF}
    \ln \sangle{ \re^{\mathbf{l}\boldsymbol{\cdot} \mathbf{v}_{2}} } \approx l_{x}\overline{v}_{2}+\frac{1}{2!}(l^{2}_{x}\sigma^{2}_{x}+l^{2}_{y}\sigma^{2}_{y})+\frac{1}{3!}(l^{3}_{x}s_{30}+3l_{x}l^{2}_{y}s_{12})\\
    +\frac{1}{4!}(l^{4}_{x}\kappa_{40}+6l^{2}_{x}l^{2}_{y}\kappa_{22}+l^{4}_{y}\kappa_{04}) 
      +\frac{1}{5!}(l^{5}_{x}p_{50}+10l^{3}_{x}l^{2}_{y}p_{32}+5l_{x}l^{4}_{y}p_{14}).
\end{multline}
One can follow the procedure for expanding the generating function of the $v_{2}\{2k\}$ cumulants~\cite{Giacalone:2016eyu} to derive expressions for the hydrodynamic probes that include higher-order moments of the \ve distribution. Instead of using Eq.~(7) from Ref.~\cite{Giacalone:2016eyu}, in this analysis  Eq.~(\ref{expGF}) is used. The difference is that in the earlier work the expansion is performed up to the skewness, while the current analysis additionally covers the 4th and 5th moments. Then, for the higher-order $v_{2}\{2k\}$ ($k=2,\dots,5$) values, we find
\begin{multline}
  \label{expand1}
    v_{2}\{4\} \approx \overline{v}_{2} + \frac{\sigma^{2}_{y}-\sigma^{2}_{x}}{2\overline{v}_{2}} - \frac{s_{30}+s_{12}}{\overline{v}_{2}^{2}} \\
    - \frac{\kappa_{40}+2\kappa_{22}+\kappa_{04}}{4\overline{v}_{2}^{3}} - \frac{5(\sigma^{2}_{y}-\sigma^{2}_{x})^{2}}{8\overline{v}_{2}^{3}} 
    + \frac{(\sigma^{2}_{y}-\sigma^{2}_{x})(3s_{30}+3s_{12})}{2\overline{v}_{2}^{4}},
\end{multline}
\begin{multline}
  \label{expand2}
    v_{2}\{6\} \approx \overline{v}_{2} + \frac{\sigma^{2}_{y}-\sigma^{2}_{x}}{2\overline{v}_{2}} - \frac{\frac{2}{3}s_{30}+s_{12}}{\overline{v}_{2}^{2}} \\
    + \frac{\kappa_{40}-\kappa_{04}}{4\overline{v}_{2}^{3}} - \frac{5(\sigma^{2}_{y}-\sigma^{2}_{x})^{2}}{8\overline{v}_{2}^{3}} + \frac{p_{50}+2p_{32}+p_{14}}{4\overline{v}_{2}^{4}} 
    + \frac{(\sigma^{2}_{y}-\sigma^{2}_{x})(4s_{30}+15s_{12})}{6\overline{v}_{2}^{4}},
\end{multline}
\begin{multline}
  \label{expand3}
    v_{2}\{8\} \approx \overline{v}_{2} + \frac{\sigma^{2}_{y}-\sigma^{2}_{x}}{2\overline{v}_{2}} - \frac{\frac{7}{11}s_{30}+s_{12}}{\overline{v}_{2}^{2}} + \frac{\frac{31}{33}\kappa_{40}+\frac{2}{11}\kappa_{22}-\kappa_{04}}{4\overline{v}_{2}^{3}} - \frac{5(\sigma^{2}_{y}-\sigma^{2}_{x})^{2}}{8\overline{v}_{2}^{3}} \\
    + \frac{\frac{5}{3}p_{50}+\frac{14}{3}p_{32}+3p_{14}}{11\overline{v}_{2}^{4}} + \frac{(\sigma^{2}_{y}-\sigma^{2}_{x})(13s_{30}+57s_{12})}{22\overline{v}_{2}^{4}},  \quad \text{and}
\end{multline}
\begin{multline}
  \label{expand4}
    v_{2}\{10\} \approx \overline{v}_{2} + \frac{\sigma^{2}_{y}-\sigma^{2}_{x}}{2\overline{v}_{2}} - \frac{\frac{12}{19}s_{30}+s_{12}}{\overline{v}_{2}^{2}} + \frac{\frac{53}{57}\kappa_{40}+\frac{4}{19}\kappa_{22}-\kappa_{04}}{4\overline{v}_{2}^{3}} - \frac{5(\sigma^{2}_{y}-\sigma^{2}_{x})^{2}}{8\overline{v}_{2}^{3}} \\
    + \frac{\frac{163}{60}p_{50}+\frac{47}{6}p_{32}+\frac{21}{4}p_{14}}{19\overline{v}_{2}^{4}} + \frac{(\sigma^{2}_{y}-\sigma^{2}_{x})(11s_{30}+\frac{99}{2}s_{12})}{19\overline{v}_{2}^{4}}.
\end{multline}
If one neglects the 5th-order moments $p$, as well as the terms with the multiplier $(\sigma^{2}_{y}-\sigma^{2}_{x})$, then the hydrodynamic probe proposed with Eq.~(14) in Ref.~\cite{Giacalone:2016eyu} is given as
\begin{equation}
\label{momHydro1}
\frac{v_{2}\{6\}-v_{2}\{8\}}{{v}_{2}\{4\}-v_{2}\{6\}} \approx \frac{1}{11} - \frac{4\kappa_{40}}{11[2\overline{v}_{2}s_{30}+3(\kappa_{40}+\kappa_{22})]}.
\end{equation}
Here, one can see that neglecting the kurtosis $\kappa$ moment, with respect to the skewness, would reduce the right-hand side of Eq.~(\ref{momHydro1}) to a constant value of 1/11. Appendix~\ref{sec:appC} shows the relation that corresponds to Eq.~(\ref{momHydro1}) if none of the terms  in Eqs.~(\ref{expand1})--(\ref{expand3}) are neglected.

The approximation given by Eq.~(\ref{momHydro1}) can be tested by expressing the right-hand side in terms of the measurable quantities ${v}_{2}\{2k\}$, with $k=2,3$, and 4. Based on Taylor expansion of the $v_{2}\{2k\}^{2}$ values, as shown in Appendix~\ref{sec:appB}, one finds
\begin{equation}
  \label{corrHydro1}
    h_{1} = \frac{v_{2}\{6\}-v_{2}\{8\}}{{v}_{2}\{4\}-v_{2}\{6\}} \approx h_{1}^{\text{Taylor}} = \frac{1}{11} - \frac{1}{11}\frac{v_{2}\{4\}^{2}-12v_{2}\{6\}^{2}+11v_{2}\{8\}^{2}}{v_{2}\{4\}^{2}-v_{2}\{6\}^{2}}
\end{equation}
if terms with $(\sigma^{2}_{y}-\sigma^{2}_{x})$ are neglected.

In the above relation, $h_{1}$ denotes the hydrodynamic probe and $h_{1}^{\text{Taylor}}$ denotes the corresponding Taylor expansion of this probe as expressed in terms of the measured $v_{2}\{2k\}$ values.

By employing the $v_{2}\{10\}$ harmonic, we define a new hydrodynamic probe $h_{2}$ as
\begin{multline}
  \label{newHydroFull}
    \frac{v_{2}\{8\}-v_{2}\{10\}}{{v}_{2}\{6\}-v_{2}\{8\}} \approx \frac{3}{19} \\
    + \frac{88p_{50}}{95[4s_{30}\overline{v}^{2}_{2}-2\overline{v}_{2}(\kappa_{40}-3\kappa_{22})-(13p_{50}+10p_{32}-3p_{14})-2(\sigma^{2}_{y}-\sigma^{2}_{x})(5s_{30}-6s_{12})]}.
  \end{multline}
In the derivation of Eq.~(\ref{newHydroFull}), none of the terms in Eqs.~(\ref{expand2})--(\ref{expand4}) have been neglected. If one neglects the 5th $p$ moment, then the right-hand side of the Eq.~(\ref{newHydroFull}) will reduce to the constant value of 3/19. A deviation from this constant value indicates that the $p$ moment cannot be neglected.

Again, as in the case with the derivation of the Eq.~(\ref{corrHydro1}), one can express the right-hand side of Eq.~(\ref{newHydroFull}) through the second power of the higher-order $v_{2}\{2k\}$ terms, with $k=3,4,5$ (as shown in Appendix~\ref{sec:appB}):
  \begin{equation}
  \label{newCorrHydroFull}
    \frac{v_{2}\{8\}-v_{2}\{10\}}{{v}_{2}\{6\}-v_{2}\{8\}} \approx \frac{3}{19} - \frac{1}{19}\frac{3v_{2}\{6\}^{2}-22v_{2}\{8\}^{2}+19v_{2}\{10\}^{2}}{v_{2}\{6\}^{2}-v_{2}\{8\}^{2}+\frac{(\sigma^{2}_{y}-\sigma^{2}_{x})s_{30}}{33\overline{v}^{3}_{2}}}.
\end{equation}
If one neglects the term with the multiplier $(\sigma^{2}_{y}-\sigma^{2}_{x})$, Eq.~(\ref{newCorrHydroFull}) reduces to
  \begin{equation}
  \label{newCorrHydro}
    h_{2} = \frac{v_{2}\{8\}-v_{2}\{10\}}{{v}_{2}\{6\}-v_{2}\{8\}} \approx h_{2}^{\text{Taylor}} = \frac{3}{19} - \frac{1}{19}\frac{3v_{2}\{6\}^{2}-22v_{2}\{8\}^{2}+19v_{2}\{10\}^{2}}{v_{2}\{6\}^{2}-v_{2}\{8\}^{2}},
\end{equation}
where $h_{2}$ denotes the new hydrodynamic probe and $h_{2}^{\text{Taylor}}$ denotes the Taylor expansion of this probe, as expressed in terms of the measured $v_{2}\{2k\}$ values.

In Section~\ref{res}, it will be shown that the term proportional to the asymmetry of the fluctuations $(\sigma^{2}_{y}-\sigma^{2}_{x})$ is negligible. The same conclusion is found in Ref.~\cite{Giacalone:2016eyu}.
 
\section{The standardized and corrected moments}
\label{moments}
  
The standardized skewness, $\gamma_{1}$, and standardized kurtosis, $\gamma_{2}$, are defined as
\begin{equation}
\label{SkewKurtTh}
\gamma_{1} \equiv \frac{s_{30}}{\sigma^{3}_{x}}, \quad \text{and}  \quad \gamma_{2} \equiv \frac{\kappa_{40}}{\sigma^{4}_{x}}.
\end{equation}
The corresponding expressions given in terms of the measured \vk values, are derived in Ref.~\cite{Bhalerao:2018anl} as
   \begin{equation}
  \label{SkewKurtv22k}
  \begin{aligned}
     \gamma^{\text{exp}}_{1} &= -2^{3/2}\frac{v_{2}\{4\}^{3}-v_{2}\{6\}^{3}}{(v_{2}\{2\}^{2}-v_{2}\{4\}^{2})^{3/2}}, \quad \text{and} \\
     \gamma^{\text{exp}}_{2} &= -\frac{3}{2}\frac{v_{2}\{4\}^{4}-12v_{2}\{6\}^{4}+11v_{2}\{8\}^{4}}{(v_{2}\{2\}^{2}-v_{2}\{4\}^{2})^{2}}.
    \end{aligned}
\end{equation} 
The standardized superskewness, $\gamma_{3}$, is defined as
\begin{equation}
\label{SkewTh}
\gamma_{3} \equiv \frac{p_{50}}{\sigma^{5}_{x}}.
\end{equation}
In order to express the standardized superskewness through the measured \vk values with $k=1,\dots,5$, one can use the expansion of the fifth power of the \vk value (as discussed in Appendix~\ref{sec:appB}). The resulting expression for the standardized superskewness, expressed in terms of measurable quantities, is
\begin{equation}
\label{Skew}
\gamma^{\text{exp}}_{3} = 6\sqrt{2}\frac{3v_{2}\{6\}^{5}-22v_{2}\{8\}^{5}+19v_{2}\{10\}^{5}}{(v_{2}\{2\}^{2}-v_{2}\{4\}^{2})^{5/2}}.
\end{equation}
The standardized moments, $\gamma^{\text{exp}}_{i}$ ($i=1,2,3$), have contributions from higher-order moments of the \ve distribution that are not negligible (as shown by Eq.~(13) in Ref.~\cite{Bhalerao:2018anl}). For example, if higher-order moments are considered, then for the standardized skewness we obtain additional terms $O_{\mathrm{N}}$ and $O_{\mathrm{D}}$, with
  \begin{equation}
    \label{gamma1ONOD}
    \gamma^{\text{exp}}_{1} = -2^{3/2}\frac{v_{2}\{4\}^{3}-v_{2}\{6\}^{3}}{(v_{2}\{2\}^{2}-v_{2}\{4\}^{2})^{3/2}} \approx -2^{3/2}\frac{-s_{30}-O_{\mathrm{N}}}{(2\sigma^{2}_{x}+O_{\mathrm{D}})^{3/2}},
  \end{equation}
  where
   \begin{equation}
  \label{SkewONOD}
  \begin{aligned}
    O_{\mathrm{N}} &= \frac{3(\kappa_{40}+\kappa_{22})}{2\overline{v}_{2}}-\frac{3(p_{50}+2p_{32}+p_{14})}{4\overline{v}^{2}_{2}}+\frac{3(\sigma^{2}_{y}-\sigma^{2}_{x})(s_{30}-2s_{12})}{2\overline{v}^{2}_{2}}+\dots, \quad \text{and} \\
O_{\mathrm{D}} &= \frac{2}{\overline{v}_{2}}(s_{30}+s_{12})+\frac{\kappa_{40}+2\kappa_{22}+\kappa_{04}}{2\overline{v}^{2}_{2}}+\frac{(\sigma^{2}_{y}-\sigma^{2}_{x})^{2}}{\overline{v}^{2}_{2}}-\frac{2(\sigma^{2}_{y}-\sigma^{2}_{x})(s_{30}-s_{12})}{\overline{v}^{3}_{2}}+\dots.
    \end{aligned}
\end{equation} 
Corrected moments, which are free of contributions from higher-order moments, can be defined if
  \begin{equation}
  \label{s12}
    s_{12} \approx \frac{s_{30}}{3}, \; \kappa_{22} \approx \frac{\kappa_{40}}{3}, \; \text{and}  \; p_{32} \approx p_{14} \approx \frac{p_{50}}{5}.
  \end{equation}
These conditions are satisfied with the elliptic power distribution~\cite{Yan:2014afa} if the eccentricity parameter $\epsilon_{0}$ in this model satisfies the condition $\epsilon_{0} \le 0.15$ and a linear flow response is assumed. Mathematical details that support the validity of the conditions given by Eq.~(\ref{s12}) are presented in Appendix~\ref{sec:appD}. With these conditions, the corrected moments can be expressed in terms of the measured \vnk ($k=1,\dots,5$) values. As an example, the corrected skewness can be expressed as
  \begin{equation}
  \label{SkewC}
\gamma^{\text{exp}}_{1\text{, corr}} = -2^{3/2} \frac{187v_{2}\{8\}^{3}-16v_{2}\{6\}^{3}-171v_{2}\{10\}^{3}}{(v_{2}\{2\}^{2}-40v_{2}\{6\}^{2}+495v_{2}\{8\}^{2}-456v_{2}\{10\}^{2})^{3/2}}.
  \end{equation}
If one replaces the powers that appear in Eq.~(\ref{SkewC}) with the expansions given in Appendix~\ref{sec:appB}, one finds
   \begin{equation}
  \label{cleanCond}
\gamma^{\text{exp}}_{1\text{, corr}} \approx -2^{3/2}\frac{-s_{30}-\overline{O}_{\mathrm{N}}}{(2\sigma^{2}_{x}+\overline{O}_{\mathrm{D}})^{3/2}} \approx \frac{s_{30}}{\sigma^{3}_{x}} \equiv \gamma_{1},
  \end{equation} 
   where $\overline{O}_{\mathrm{N}}$ and $\overline{O}_{\mathrm{D}}$ are given by
   \begin{equation}
     \label{SkewObarNObarD}
     \begin{aligned}
       \overline{O}_{\mathrm{N}} &= \frac{3(\sigma^{2}_{y}-\sigma^{2}_{x})(s_{30}-2s_{12})}{2\overline{v}^{2}_{2}}+\dots, \quad \text{and} \\
       \overline{O}_{\mathrm{D}} &= -\frac{2(\sigma^{2}_{y}-\sigma^{2}_{x})(s_{30}-s_{12})}{\overline{v}^{3}_{2}}+\dots.
     \end{aligned}
   \end{equation}
We note that the contribution from the term proportional to $(\sigma^{2}_{y}-\sigma^{2}_{x})$ as well as the contributions from moments of order higher than 5 are negligible.

The same correction procedure can also be applied to the standardized kurtosis and the standardized superskewness. The corrected form of these moments are given by
  \begin{equation}
  \label{kurtosisC}
  \begin{aligned}  
    \gamma^{\text{exp}}_{2\text{, corr}} &= -\frac{3}{2} \frac{v_{2}\{4\}^{4}+24v_{2}\{6\}^{4}-253v_{2}\{8\}^{4}+228v_{2}\{10\}^{4}}{(v_{2}\{2\}^{2}-40v_{2}\{6\}^{2}+495v_{2}\{8\}^{2}-456v_{2}\{10\}^{2})^{2}},\quad \text{and}\\
    \gamma^{\text{exp}}_{3\text{, corr}} &= 6\sqrt{2} \frac{3v_{2}\{6\}^{5}-22v_{2}\{8\}^{5}+19v_{2}\{10\}^{5}}{(v_{2}\{2\}^{2}-40v_{2}\{6\}^{2}+495v_{2}\{8\}^{2}-456v_{2}\{10\}^{2})^{5/2}}.
  \end{aligned}
\end{equation}

\section{CMS detector, event selection, and systematic uncertainties}
\label{expData}

The central feature of the CMS apparatus is a superconducting solenoid of 6\unit{m} internal diameter, providing a magnetic field of 3.8\unit{T}. Within the solenoid volume are a silicon pixel and strip tracker, a lead tungstate crystal electromagnetic calorimeter (ECAL), and a brass and scintillator hadron calorimeter (HCAL), each composed of a barrel and two endcap sections. Forward calorimeters (HF), made of steel and quartz-fibres, extend the pseudorapidity ($\eta$) coverage provided by the barrel and endcap detectors. Muons are measured in gas-ionization detectors embedded in the steel flux-return yoke outside the solenoid. A more detailed description of the CMS detector, together with a definition of the coordinate system used and the relevant kinematic variables, can be found in Ref.~\cite{CMS:2008xjf}.

Events of interest are selected using a two-tiered trigger system. The first level (L1), composed of custom hardware processors, uses information from the calorimeters and muon detectors to select events at a rate of around 100\unit{kHz}~\cite{CMS:2020cmk}. The second level, known as the high-level trigger (HLT), consists of a farm of processors running a version of the full event reconstruction software optimized for fast processing, and reduces the event rate to about 5 to 8\unit{kHz} before data storage~\cite{CMS:2016ngn}. 

The data analyzed in this paper, before applying the selection described below, consist of 4.27\ten{9} minimum bias lead-lead (\PbPb) collision events at \sqrtsNN = 5.02\TeV, collected in 2018 with an integrated luminosity of 0.607\nbinv~\cite{CMS:LUM-17-003,CMS:LUM-18-001}. The minimum bias events are triggered by requiring signals above readout thresholds of 3\GeV in each of the two HF calorimeters~\cite{CMS:2016ngn}. Further selections are applied to reject events from beam-gas interactions and nonhadronic collisions~\cite{CMS:2016xef}. The events are also required to have at least one reconstructed primary vertex (based on two or more tracks) within a distance ($\abs{z_{\text{vtx}}}$) of 15\unit{cm} from the nominal interaction point along the beam axis. If multiple vertices are found in an event then the primary vertex is selected as the one with the highest track multiplicity in the event. The total energy deposited in both HF calorimeters is used for determination of the event centrality~\cite{CMS:2011iwn}.

Only tracks that satisfy the high-purity selection criteria described in Ref.~\cite{CMS:2016xef} are used in the analysis. In addition, a reconstructed track is only considered as a candidate track from the primary vertex if the separation along the beam axis between the track and the vertex ($d_{\mathrm{z}}$), and the track-vertex impact parameter measured transverse to the beam ($d_{\mathrm{xy}}$), are both less than three times their respective uncertainties $\sigma_{\mathrm{z}}$ and $\sigma_{\mathrm{xy}}$. The relative uncertainty in the transverse momentum ($\sigma_{\pt}/$\pt) measurement is required to be less than 10\%. In addition, tracks must have 11 or more hits ($n_{\text{hits}}$) along their trajectory in the pixel and strip tracking detectors combined. The relative $\chi^{2}$ of the track fit ($\chi^{2}/N_{\mathrm{dof}}/N_{\text{layers}}$) is required to be less than 0.18, where $N_{\mathrm{dof}}$ is the number of degrees of freedom during the track fitting and $N_{\text{layers}}$ is the number of tracking layers used.

Each reconstructed track is weighted by the inverse of the efficiency factor $w(\eta, \pt, \text{centrality})$ calculated based on Monte Carlo \HYDJET 1.9~\cite{Lokhtin:2005px}. The efficiency factor accounts for the reconstruction efficiency $\epsilon$ and the fraction of misreconstructed tracks $f_{\text{misrec}}(\eta,\pt,\text{centrality})$, with $w=\epsilon/(1-f_{\text{misrec}})$~\cite{CMS:2014pgm}. The response of the CMS detector to the generated events is simulated with \GEANTfour~\cite{GEANT4:2002zbu}.

The systematic uncertainties are determined by varying the vertex selection, the track selection, the centrality determination, and the efficiency correction. For each of the indicated sources of systematic uncertainty, an absolute difference between results from the varied and nominal values are used to obtain the uncertainty. The vertex selection uncertainty is found by performing the analysis with $\abs{z_{\mathrm{vtx}}} < 3$\unit{cm} and with $3 < \abs{z_{\mathrm{vtx}}} < 15$\unit{cm}. The uncertainty related to the track selection is determined by comparing tight ($d_{\mathrm{z}}/\sigma_{\mathrm{z}} < 2$, $d_{\mathrm{xy}}/\sigma_{\mathrm{xy}} < 2$, $\sigma_{\pt}/\pt < 0.05$ and $\chi^{2}/N_{\mathrm{dof}}/N_{\text{layers}} < 0.15$) criterion and loose ($d_{\mathrm{z}}/\sigma_{\mathrm{z}} < 5$, $d_{\mathrm{xy}}/\sigma_{\mathrm{xy}} < 5$, $\sigma_{\pt}/\pt < 0.15$ and $\chi^{2}/N_{\mathrm{dof}}/N_{\text{layers}} < 0.18$) criterion for tracks with the results obtained with the nominal values. The finite HF energy resolution results in small migration of events across the centrality bin boundaries. The centrality calibration is varied to calculate the effect of the minimum bias event selection efficiency of the HF calorimeters~\cite{CMS:2021kvd}. Differences in the \pt spectra between the data and the Monte Carlo simulation can affect the efficiency factor. To estimate the systematic uncertainty arising from these differences, the efficiency factor $w$ is varied by $\pm$~5\%. The final systematic uncertainty is taken as a quadratic sum of the four indicated sources, with the track selection found to have the dominant contribution.

\section{Results}
\label{res}

\begin{figure}[ht!]
     \centering
  \includegraphics[width=0.65\textwidth]{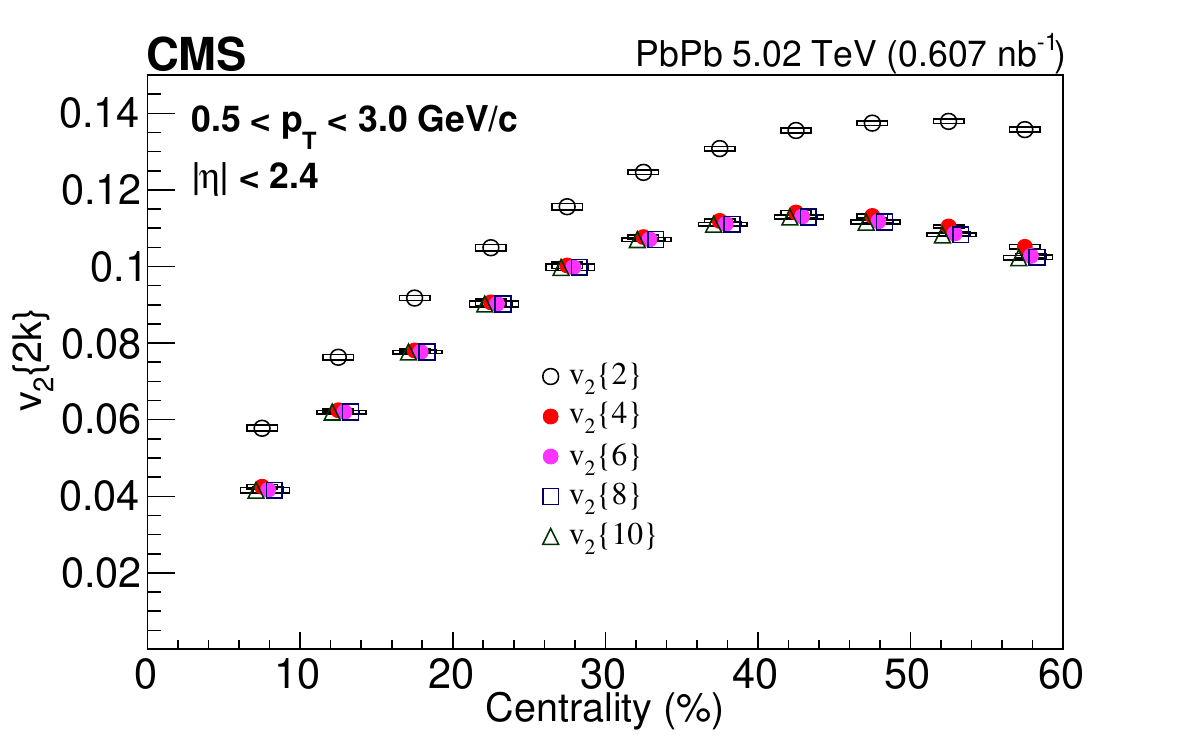}
  \caption{\label{fig:2} The $v_{2}\{2k\}$ ($k = 1,\dots,5$) values as functions of centrality in \PbPb collisions at \sqrtsNN = 5.02\TeV. The vertical sizes of the open boxes denote the systematic uncertainties. Statistical uncertainties are negligible compared to the marker size. Points are plotted at the center of the respective centrality ranges. The markers are displaced horizontally for better visibility.}
\end{figure}

It was previously shown in Ref.~\cite{Giacalone:2016eyu} that the presence of non-Gaussian fluctuations in the initial-state energy density leads to a fine splitting between the higher-order \vnk cumulants. The \vk ($k = 1,\dots,5$) values are presented in Fig.~\ref{fig:2} as functions of centrality in \PbPb collisions at \sqrtsNN = 5.02\TeV. These results are obtained from charged particles with $\abs{\eta} < 2.4$ and $0.5 < \pt < 3.0$\GeVc. The statistical uncertainties are about two orders of magnitude smaller than the systematic ones. A clear splitting between the \vc{2} and higher-orders cumulant based \vk ($k = 2,\dots,5$) values is visible. The difference is attributed to flow fluctuations, with $v_{2}\{2\}^{2} \approx v_{2}\{2k\}^{2} + 2\sigma^{2}_{\mathrm{v}}$ for $k > 1$~\cite{Giacalone:2016eyu,CMS:2022bmk}, where $\sigma^{2}_{\mathrm{v}}$ is the \ve variance. From Fig.~\ref{fig:2}, it is clear that the flow fluctuations become larger going to more peripheral collisions (higher centrality percentages).

\begin{figure}[ht!]
     \centering
  \includegraphics[width=0.65\textwidth]{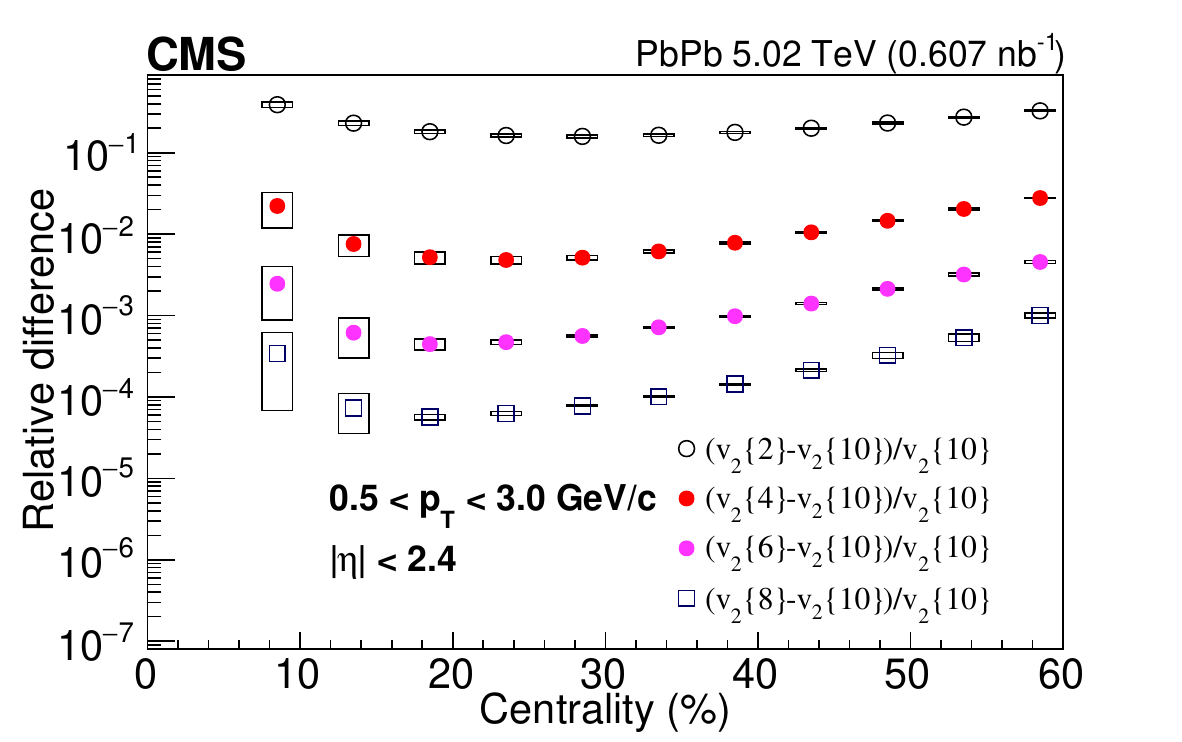}
  \caption{\label{fig:3} The relative differences $(v_{2}\{2k\}-v_{2}\{10\})/v_{2}\{10\}$ ($k = 1,\dots,4$) as functions of centrality in \PbPb collisions at $\sqrtsNN = 5.02\TeV$. The vertical sizes of the open boxes denote the systematic uncertainties. Statistical uncertainties are negligible compared to the marker size. Points are plotted at the center of the respective centrality ranges.}
\end{figure}

In Fig.~\ref{fig:2}, the ordering and a fine splitting between the $v_{2}\{2k\}$ ($k = 2, \dots, 5$) values are not clearly visible. The \vk ($k = 1,\dots 5$) values need to satisfy the ordering relationship $v_{2}\{2k\} > v_{2}\{2(k+1)\}$ in order to define the hydrodynamic probes given by Eqs.~(\ref{corrHydro1}) and (\ref{newCorrHydro}). The splitting between the higher-order cumulant based $v_{2}\{2k\}$ values $v_{2}\{4\} \gtrsim v_{2}\{6\} \gtrsim v_{2}\{8\} \gtrsim v_{2}\{10\}$ becomes much finer than the splitting with respect to the lowest-order cumulant based $v_{2}\{2\}$ value. Figure~\ref{fig:3} shows the corresponding relative differences $(v_{2}\{2k\}-v_{2}\{10\})/v_{2}\{10\}$ ($k = 1,\dots,4$) as functions of centrality. The magnitudes of these differences are plotted using a logarithmic scale. The fine splitting between cumulants of different orders as well as the expected hydrodynamic ordering of the splittings is clearly seen. The relative difference between adjacent $v_{2}\{2k\}-v_{2}\{10\}$ values decreases by about an order of magnitude for each increment in $k$.

\begin{figure}[ht!]
     \centering
  \includegraphics[width=0.65\textwidth]{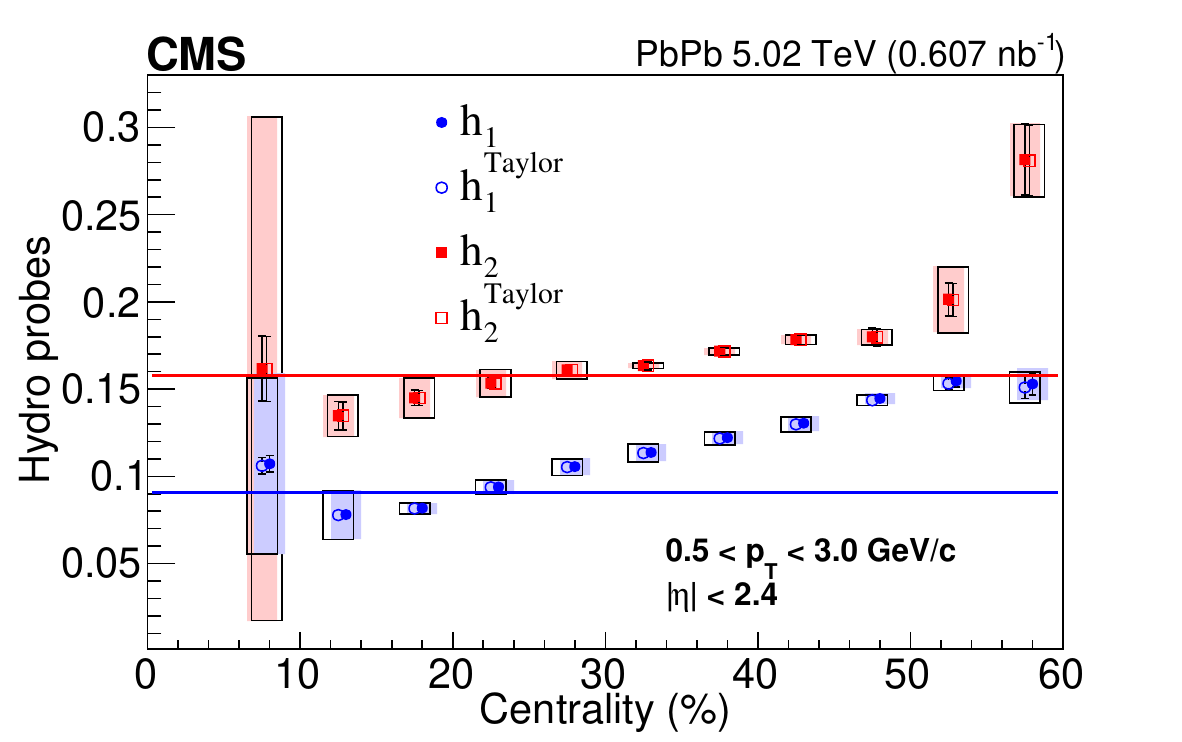}
  \caption{\label{fig:4} The $h_{1}=(v_{2}\{6\}-v_{2}\{8\})/({v}_{2}\{4\}-v_{2}\{6\})$ (closed blue circles) and the $h_{2}=(v_{2}\{8\}-v_{2}\{10\})/({v}_{2}\{6\}-v_{2}\{8\})$ (closed red squares) hydrodynamic probes as functions of centrality in \PbPb collisions at \sqrtsNN = 5.02\TeV. The distributions depicted with the open circles and squares represent the corresponding Taylor expansions given by $h_{1}^{\text{Taylor}}$ and $h_{2}^{\text{Taylor}}$, respectively. The horizontal blue (red) line represents a constant value of 1/11 (3/19). The bars (the vertical sizes of the open boxes) denote the statistical (systematic) uncertainties. Points are plotted at the center of the respective centrality ranges.}
\end{figure}

The measured cumulants of different orders are used to calculate the hydrodynamic probes given by the left-hand sides of Eqs.~(\ref{corrHydro1}) and (\ref{newCorrHydro}). Figure~\ref{fig:4} displays these distributions with closed symbols. Open symbols in the same figure show the right-hand sides of Eqs.~(\ref{corrHydro1}) and (\ref{newCorrHydro}). These are also constructed using the measured $v_{2}\{2k\}$ ($k = 1, \dots, 5$) cumulants.

As a consequence of the low multiplicities involved in the calculation of the cumulant based \vk values, statistical and systematic uncertainties quickly increase as peripheral collisions are approached. Also, in the region of very central collisions, because of the small magnitudes of the cumulant based $v_{2}\{2k\}$ values, the statistical and systematic uncertainties are larger and, therefore, statistical and systematic uncertainties of the hydrodynamic probes are increased. The magnitudes of the statistical uncertainties of the $h_{2}$ distribution are larger with respect to those corresponding to the $h_{1}$ distribution because higher-order cumulants are involved. After the 0--5\% centrality bin, where the measurement uncertainties are large, the $h_{1}$ and $h_{2}$ values show an increasing trend going to more peripheral events. Based on an event-by-event measurement of the \ve distribution, it was reported in Ref.~\cite{CMS:2017glf} that the $h_{1}$ has a value of $0.143 \pm 0.008\stat \pm 0.014\syst$ for 20--25\% central events, and  increases to $0.185 \pm 0.005\stat \pm 0.012\syst$ as the centrality increases to 55--60\%. Using the Q-cumulant method, Ref.~\cite{ALICE:2018rtz} compares the measured $v_{2}\{6\}-v_{2}\{8\}$ and $\frac{1}{11}(v_{2}\{4\}-v_{2}\{6\})$ distributions. Taking into consideration differences in acceptance, the current analysis is consistent with the results of previous analyses~\cite{CMS:2017glf,ALICE:2018rtz}, although with smaller uncertainties. There is a very good agreement between the distributions obtained by the expressions given on left-hand and right-hand sides of Eq.~(\ref{corrHydro1}) and Eq.~(\ref{newCorrHydro}) in Fig.~\ref{fig:4}. This indicates the importance of including higher-order terms in the Taylor expansion of the \ve generating function.

\begin{figure}[ht!]
  \centering
  \includegraphics[width=0.77\textwidth]{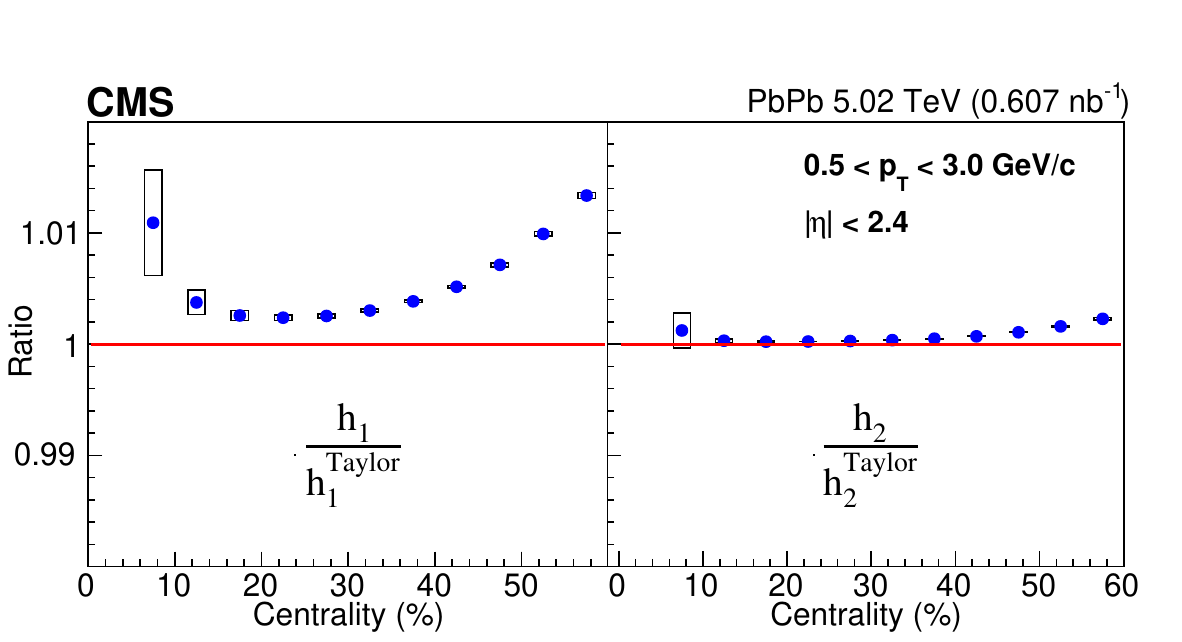}
  \caption{\label{fig:5} The ratios between the hydrodynamic probes and their Taylor expansions. The ratios are plotted as functions of centrality in \PbPb collisions at \sqrtsNN = 5.02\TeV. The vertical sizes of the open boxes denote systematic uncertainties. Statistical uncertainties are negligible compared to the marker size. Points are plotted at the center of the respective centrality ranges.}
\end{figure}

Figure~\ref{fig:5} shows ratios between the left- and right-hand sides of the hydrodynamic probes given by Eq.~(\ref{corrHydro1}) (left plot) and Eq.~(\ref{newCorrHydro}) (right plot) as functions of centrality. The two sides  differ by less than 1.3\% and 0.3\% in the cases of Eqs.~(\ref{corrHydro1}) and (\ref{newCorrHydro}), respectively. Thus, based on Eqs.~(\ref{corrHydro1Full}) and (\ref{newCorrHydroFull}), which differ from the Eqs.~(\ref{corrHydro1}) and (\ref{newCorrHydro}) only by the term that is proportional to $(\sigma^{2}_{y}-\sigma^{2}_{x})$, one can conclude that the contribution of this term is quite small and can be neglected.

\begin{figure}[ht!]
  \centering
  \includegraphics[width=0.65\textwidth]{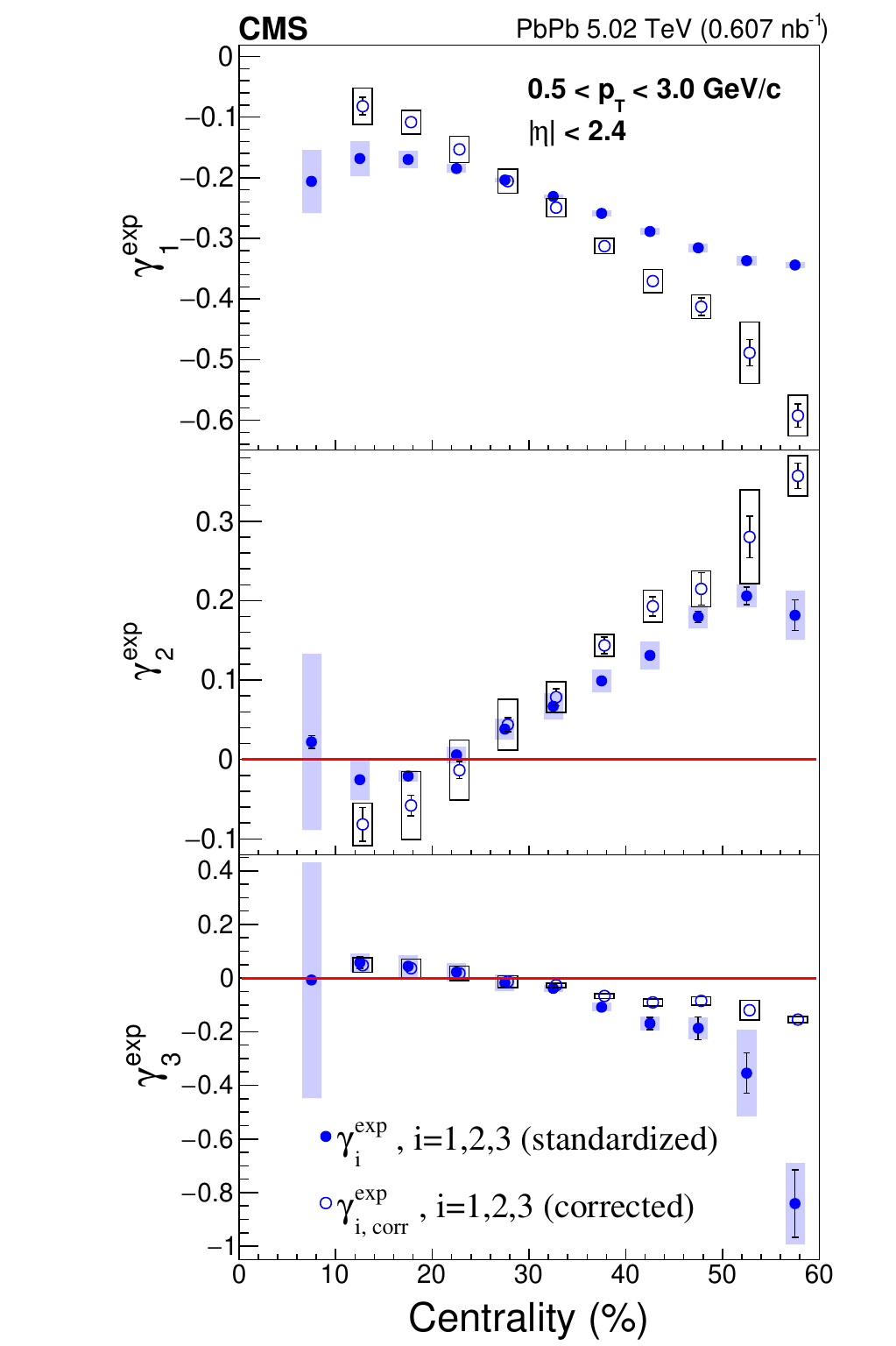}
  \caption{\label{fig:6} The magnitudes of the measured (closed circles) standardized skewness $\gamma^{\text{exp}}_{1}$ (upper), standardized kurtosis $\gamma^{\text{exp}}_{2}$ (middle), and standardized superskewness $\gamma^{\text{exp}}_{3}$ (lower) as functions of centrality in \PbPb collisions at $\sqrtsNN = 5.02\TeV$. The magnitudes of the corrected skewness $\gamma^{\text{exp}}_{1\text{ ,corr}}$ (upper), corrected kurtosis $\gamma^{\text{exp}}_{2\text{, corr}}$ (middle), and corrected superskewness $\gamma^{\text{exp}}_{3\text{, corr}}$ (lower) are presented with the open circles. The bars (the vertical sizes of the open boxes) denote statistical (systematic) uncertainties. Points are plotted at the center of the respective centrality ranges.}
\end{figure}

Figure~\ref{fig:6} displays the distributions of the measured standardized skewness $\gamma^{\text{exp}}_{1}$, standardized kurtosis $\gamma^{\text{exp}}_{2}$, and standardized superskewness $\gamma^{\text{exp}}_{3}$ as functions of centrality. The distributions of the corrected skewness $\gamma^{\text{exp}}_{1\text{, corr}}$ (upper), corrected kurtosis $\gamma^{\text{exp}}_{2\text{, corr}}$ (middle), and corrected superskewness $\gamma^{\text{exp}}_{3\text{, corr}}$ (lower) are presented with the open blue circles. The systematic uncertainties are generally larger than the statistical ones. The values of the corrected moments in the first analyzed centrality bin (5--10\%) are not shown because of the very large systematic uncertainties for this centrality range.

Although with different acceptance and differences in the applied methodology, the obtained results for the experimentally measured standardized skewness are in a fair agreement with the results obtained in Ref. \cite{CMS:2017glf}. The $\gamma^{\text{exp}}_{1}$ is negative over the entire analyzed centrality range. This is a consequence of the $v_{2,x}$ distribution having a long tail on the low $v_{2,x}$ side (as shown in Fig.~1 of Ref.~\cite{Giacalone:2016eyu}). The $v_{2,y}$ distribution is symmetric and thus its skewness is equal to zero. The $\gamma^{\text{exp}}_{1}$ had been predicted to become more negative as the centrality percentile increases~\cite{Giacalone:2016eyu}. This measurement confirms the prediction.

Similarly to the case of the skewness, although with different acceptance and a difference in the applied methodology to obtain the \vk values, the  $\gamma^{\text{exp}}_{2}$ values are in reasonable agreement with the results presented in Ref.~\cite{Bhalerao:2018anl}. The $\gamma^{\text{exp}}_{2}$ values are positive over nearly the entire analyzed centrality range. The exception is the centrality range of 10--20\%. The sign of $\gamma^{\text{exp}}_{2}$ is driven by the mean eccentricity $\epsilon_{0}$, and is predicted to be negative for $\epsilon_{0} < 0.28$ and positive for $\epsilon_{0} > 0.29$~\cite{Bhalerao:2018anl}. Our results are qualitatively in an agreement with this prediction. If the \ve values were to reflect a purely linear response to the initial-state eccentricity $\epsilon_{2}$, the corresponding kurtosis would equal the kurtosis of the initial $\epsilon_{2}$ fluctuations.  Although a fully linear response is not expected, calculations predict that deviations from a Gaussian behavior for the skewness and kurtosis, although significantly reduced by the hydrodynamic evolution, are still influenced by the early stage hydrodynamics~\cite{Giacalone:2016eyu,Abbasi:2017ajp}.

Except for collisions with centrality less than 25\%, where it is either positive or vanishes, the $\gamma^{\text{exp}}_{3}$ moment is negative, with its absolute magnitude increasing towards more peripheral collisions. This is the first measurement of this moment of the \ve distribution. Without the superskewness, it would be impossible to describe the centrality dependence observed for the new hydrodynamic probe $h_{2}$.

In addition to the experimental results for the standardized skewness, kurtosis, and superskewness, the corresponding corrected moments are also presented. Except for the superskewness, the corrected skewness and kurtosis have larger slopes with respect to the standardized ones. The corrected moments give additional constraints on models of the initial-state geometry.

\section{Summary}
\label{sum}

The elliptic anisotropy values based on different orders of multiparticle cumulants, $v_{2}\{2k\}$ ($1 \le k \le 5$), are determined as functions of centrality in lead-lead collisions at a center-of-mass energy per nucleon pair of \sqrtsNN = 5.02\TeV, with an integrated luminosity of 0.607\nbinv. The \vc{10} value is determined for the first time. A fine splitting is observed between the harmonic values based on different cumulant orders, with $\vc{4} \gtrsim \vc{6} \gtrsim \vc{8} \gtrsim \vc{10}$. The ordering of the \vk values is consistent with a hydrodynamic evolution of the quark-gluon plasma (QGP). This splitting is attributed to a non-Gaussian behavior in the event-by-event fluctuations of the \ve distribution, leading to nonzero values of the skewness, kurtosis, and superskewness. The splitting becomes finer as the $k$ value increases, with the difference between the adjacent $v_{2}\{2k\}$ values decreasing by about an order of magnitude for each increment. The standardized magnitude of the \ve moments are presented, together with their corrected values, where contributions from higher-order moments (up to the 5th moment) are removed. The large data set of lead-lead collisions collected by the CMS experiment enables a precise measurement of the hydrodynamic probe $h_{1}$ as a function of centrality, where $h_{1}=(v_{2}\{6\}-v_{2}\{8\})/({v}_{2}\{4\}-v_{2}\{6\})$. A strong centrality dependence is observed, with values slowly increasing going to more peripheral collisions. This contrasts with an earlier hydrodynamic expectation that had taken the skewness of the initial-state geometry as the main source of non-Gaussian fluctuations. In that case, the ratio was not expected to depend on centrality. Based on the first $v_{2}\{10\}$ measurements, a new hydrodynamic probe is introduced that gives an even more precise measure of the initial-state geometry assuming a hydrodynamic evolution of the QGP. The new probe, $h_{2}$, defined as $h_{2}=(v_{2}\{8\}-v_{2}\{10\})/({v}_{2}\{6\}-v_{2}\{8\})$, is also found to have a centrality dependence, with a shape similar to the $h_{1}$ results. The centrality dependence of both ratios can be understood in terms of the evolving shape of the interaction region with centrality. Based on these results, centrality-dependent moments for the fluctuation-driven event-by-event \ve distribution are determined, including the skewness, kurtosis and, for the first time, superskewness. The results provide basic input for a precision test of models that assume a hydrodynamic expansion of the QGP.

\begin{acknowledgments}
We congratulate our colleagues in the CERN accelerator departments for the excellent performance of the LHC and thank the technical and administrative staffs at CERN and at other CMS institutes for their contributions to the success of the CMS effort. In addition, we gratefully acknowledge the computing centers and personnel of the Worldwide LHC Computing Grid and other centers for delivering so effectively the computing infrastructure essential to our analyses. Finally, we acknowledge the enduring support for the construction and operation of the LHC, the CMS detector, and the supporting computing infrastructure provided by the following funding agencies: SC (Armenia), BMBWF and FWF (Austria); FNRS and FWO (Belgium); CNPq, CAPES, FAPERJ, FAPERGS, and FAPESP (Brazil); MES and BNSF (Bulgaria); CERN; CAS, MoST, and NSFC (China); MINCIENCIAS (Colombia); MSES and CSF (Croatia); RIF (Cyprus); SENESCYT (Ecuador); MoER, ERC PUT and ERDF (Estonia); Academy of Finland, MEC, and HIP (Finland); CEA and CNRS/IN2P3 (France); SRNSF (Georgia); BMBF, DFG, and HGF (Germany); GSRI (Greece); NKFIH (Hungary); DAE and DST (India); IPM (Iran); SFI (Ireland); INFN (Italy); MSIP and NRF (Republic of Korea); MES (Latvia); LAS (Lithuania); MOE and UM (Malaysia); BUAP, CINVESTAV, CONACYT, LNS, SEP, and UASLP-FAI (Mexico); MOS (Montenegro); MBIE (New Zealand); PAEC (Pakistan); MES and NSC (Poland); FCT (Portugal); MESTD (Serbia); MCIN/AEI and PCTI (Spain); MOSTR (Sri Lanka); Swiss Funding Agencies (Switzerland); MST (Taipei); MHESI and NSTDA (Thailand); TUBITAK and TENMAK (Turkey); NASU (Ukraine); STFC (United Kingdom); DOE and NSF (USA).

\hyphenation{Rachada-pisek} Individuals have received support from the Marie-Curie program and the European Research Council and Horizon 2020 Grant, contract Nos.\ 675440, 724704, 752730, 758316, 765710, 824093, and COST Action CA16108 (European Union); the Leventis Foundation; the Alfred P.\ Sloan Foundation; the Alexander von Humboldt Foundation; the Science Committee, project no. 22rl-037 (Armenia); the Belgian Federal Science Policy Office; the Fonds pour la Formation \`a la Recherche dans l'Industrie et dans l'Agriculture (FRIA-Belgium); the Agentschap voor Innovatie door Wetenschap en Technologie (IWT-Belgium); the F.R.S.-FNRS and FWO (Belgium) under the ``Excellence of Science -- EOS" -- be.h project n.\ 30820817; the Beijing Municipal Science \& Technology Commission, No. Z191100007219010 and Fundamental Research Funds for the Central Universities (China); the Ministry of Education, Youth and Sports (MEYS) of the Czech Republic; the Shota Rustaveli National Science Foundation, grant FR-22-985 (Georgia); the Deutsche Forschungsgemeinschaft (DFG), under Germany's Excellence Strategy -- EXC 2121 ``Quantum Universe" -- 390833306, and under project number 400140256 - GRK2497; the Hellenic Foundation for Research and Innovation (HFRI), Project Number 2288 (Greece); the Hungarian Academy of Sciences, the New National Excellence Program - \'UNKP, the NKFIH research grants K 124845, K 124850, K 128713, K 128786, K 129058, K 131991, K 133046, K 138136, K 143460, K 143477, 2020-2.2.1-ED-2021-00181, and TKP2021-NKTA-64 (Hungary); the Council of Science and Industrial Research, India; ICSC -- National Research Center for High Performance Computing, Big Data and Quantum Computing, funded by the EU NexGeneration program (Italy); the Latvian Council of Science; the Ministry of Education and Science, project no. 2022/WK/14, and the National Science Center, contracts Opus 2021/41/B/ST2/01369 and 2021/43/B/ST2/01552 (Poland); the Funda\c{c}\~ao para a Ci\^encia e a Tecnologia, grant CEECIND/01334/2018 (Portugal); the National Priorities Research Program by Qatar National Research Fund; MCIN/AEI/10.13039/501100011033, ERDF ``a way of making Europe", and the Programa Estatal de Fomento de la Investigaci{\'o}n Cient{\'i}fica y T{\'e}cnica de Excelencia Mar\'{\i}a de Maeztu, grant MDM-2017-0765 and Programa Severo Ochoa del Principado de Asturias (Spain); the Chulalongkorn Academic into Its 2nd Century Project Advancement Project, and the National Science, Research and Innovation Fund via the Program Management Unit for Human Resources \& Institutional Development, Research and Innovation, grant B37G660013 (Thailand); the Kavli Foundation; the Nvidia Corporation; the SuperMicro Corporation; the Welch Foundation, contract C-1845; and the Weston Havens Foundation (USA).
\end{acknowledgments}

\bibliography{auto_generated}

\clearpage

\appendix
\numberwithin{equation}{section}
\section{Analytic calculation of \texorpdfstring{$\vten$}{vten} statistical uncertainty}
  \label{sec:appA}

The statistical uncertainty in $v_{n}\{10\}$  (given by Eq.~(\ref{FCoeff})) can be expressed analytically using the procedure described in Ref.~\cite{Nadderd:2021fst}. 
Based on the experimental  $\dangle{2m}$ azimuthal correlations, the $v_{n}\{10\}$ statistical uncertainty is given by 

\begin{equation}
  \label{DeltaCumul}
   \begin{aligned}
      s^{2}[v_{n}\{10\}]4560^{2}(v_{n}\{10\})^{18} &= A^{2}\sigma^{2}_{\dangle{2}} + B^{2}\sigma^{2}_{\dangle{4}} + C^{2}\sigma^{2}_{\dangle{6}} + D^{2}\sigma^{2}_{\dangle{8}} + \sigma^{2}_{\dangle{10}} \\
      &\quad + 2AB\sigma_{\dangle{2},\dangle{4}} + 2AC\sigma_{\dangle{2},\dangle{6}} + 2AD\sigma_{\dangle{2},\dangle{8}} \\
      &\quad + 2A\sigma_{\dangle{2},\dangle{10}} + 2BC\sigma_{\dangle{4},\dangle{6}} + 2BD\sigma_{\dangle{4},\dangle{8}} \\
      &\quad + 2B\sigma_{\dangle{4},\dangle{10}} + 2CD\sigma_{\dangle{6},\dangle{8}} + 2C\sigma_{\dangle{6},\dangle{10}} \\
      &\quad + 2D\sigma_{\dangle{8},\dangle{10}},
      \end{aligned}
\end{equation}
where $\sigma^{2}_{\dangle{2k}}$, $k=1, \dots ,5$ and $\sigma_{\dangle{2k},\dangle{2l}}$, with $k=1, \dots ,4$ and $l=2, \dots ,5$ are corresponding variances and covariances, respectively. The coefficients $A$, $B$, $C$, and $D$ are given by
\begin{equation}
  \label{ABC}
  \begin{aligned}
    A &= 14400\dangle{2}^{4} - 10800 \dangle{2}^{2} \dangle{4} + 800 \dangle{6} \dangle{2} + 900 \dangle{4}^{2} - 25 \dangle{8},\\
    B &= 1800 \dangle{4} \dangle{2} - 3600\dangle{2}^{3} - 100\dangle{6},\\
    C &= 400\dangle{2}^{2} - 100\dangle{4},\quad\text{and}\\
    D &= -25\dangle{2}.
      \end{aligned}
\end{equation}

\section{Central moment expansions of \texorpdfstring{$\vk^2$}{vk2}, \texorpdfstring{$\vk^3$}{vk3}, \texorpdfstring{$\vk^4$}{vk4}, and \texorpdfstring{$\vk^5$}{vk5}}
\label{sec:appB}

To obtain expressions for the hydrodynamic probes and the standardized and corrected moments in terms of the measured $\vk$ values, it is necessary to express different powers of the cumulant \vk values in terms of the central moments. This appendix shows the needed expressions for $\vk^2$, $\vk^3$, $\vk^4$, and $\vk^5$. The expressions only extend to the fifth central moments.

For $\vk^2$:
\begin{equation}
  \label{expand21}
  \begin{aligned}
    v_{2}\{4\}^{2}&\approx \overline{v}^{2}_{2} + (\sigma^{2}_{y}-\sigma^{2}_{x}) - \frac{2s_{30}+2s_{12}}{\overline{v}_{2}} - \frac{\kappa_{40}+2\kappa_{22}+\kappa_{04}}{2\overline{v}_{2}^{2}} \\
    &\quad - \frac{(\sigma^{2}_{y}-\sigma^{2}_{x})^{2}}{\overline{v}_{2}^{2}} + \frac{2(\sigma^{2}_{y}-\sigma^{2}_{x})(s_{30}+s_{12})}{\overline{v}_{2}^{3}} ,\\
    v_{2}\{6\}^{2}&\approx \overline{v}^{2}_{2} + (\sigma^{2}_{y}-\sigma^{2}_{x}) - \frac{4s_{30}+6s_{12}}{3\overline{v}_{2}} + \frac{\kappa_{40}-\kappa_{04}}{2\overline{v}_{2}^{2}} \\
    &\quad - \frac{(\sigma^{2}_{y}-\sigma^{2}_{x})^{2}}{\overline{v}_{2}^{2}} + \frac{p_{50}+2p_{32}+p_{14}}{2\overline{v}_{2}^{3}} + \frac{(\sigma^{2}_{y}-\sigma^{2}_{x})(\frac{2}{3}s_{30}+4s_{12})}{\overline{v}_{2}^{3}} ,\\
    v_{2}\{8\}^{2}&\approx \overline{v}^{2}_{2} + (\sigma^{2}_{y}-\sigma^{2}_{x}) - \frac{14s_{30}+22s_{12}}{11\overline{v}_{2}} + \frac{\frac{31}{33}\kappa_{40}+\frac{2}{11}\kappa_{22}-\kappa_{04}}{2\overline{v}_{2}^{2}} \\
	  &\quad - \frac{(\sigma^{2}_{y}-\sigma^{2}_{x})^{2}}{\overline{v}_{2}^{2}} + \frac{10p_{50}+28p_{32}+18p_{14}}{33\overline{v}_{2}^{3}} + \frac{(\sigma^{2}_{y}-\sigma^{2}_{x})(6s_{30}+46s_{12})}{11\overline{v}_{2}^{3}},\quad\text{and}\\
    v_{2}\{10\}^{2}&\approx \overline{v}^{2}_{2} + (\sigma^{2}_{y}-\sigma^{2}_{x}) - \frac{24s_{30}+38s_{12}}{19\overline{v}_{2}} + \frac{\frac{53}{19}\kappa_{40}+\frac{12}{19}\kappa_{22}-3\kappa_{04}}{6\overline{v}_{2}^{2}} \\
    &\quad - \frac{(\sigma^{2}_{y}-\sigma^{2}_{x})^{2}}{\overline{v}_{2}^{2}} + \frac{\frac{163}{30}p_{50}+\frac{47}{3}p_{32}+\frac{21}{2}p_{14}}{19\overline{v}_{2}^{3}} + \frac{(\sigma^{2}_{y}-\sigma^{2}_{x})(10s_{30}+80s_{12})}{19\overline{v}_{2}^{3}}.
  \end{aligned}
\end{equation}
For $\vk^3$:
\begin{equation}
  \label{expand22}
  \begin{aligned}
    v_{2}\{4\}^{3}&\approx \overline{v}^{3}_{2} + \frac{3\overline{v}_{2}(\sigma^{2}_{y}-\sigma^{2}_{x})}{2} - (3s_{30}+3s_{12}) - \frac{3\kappa_{40}+6\kappa_{22}+3\kappa_{04}}{4\overline{v}_{2}} \\
    &\quad - \frac{9(\sigma^{2}_{y}-\sigma^{2}_{x})^{2}}{8\overline{v}_{2}} + \frac{3(\sigma^{2}_{y}-\sigma^{2}_{x})(s_{30}+s_{12})}{2\overline{v}_{2}^{2}} ,\\
    v_{2}\{6\}^{3}&\approx \overline{v}^{3}_{2} + \frac{3\overline{v}_{2}(\sigma^{2}_{y}-\sigma^{2}_{x})}{2} - (2s_{30}+3s_{12}) - \frac{-3\kappa_{40}+3\kappa_{04}}{4\overline{v}_{2}} \\
    &\quad - \frac{9(\sigma^{2}_{y}-\sigma^{2}_{x})^{2}}{8\overline{v}_{2}} + \frac{3p_{50}+6p_{32}+3p_{14}}{4\overline{v}_{2}^{2}} + \frac{9(\sigma^{2}_{y}-\sigma^{2}_{x})s_{12}}{2\overline{v}_{2}^{2}} ,\\
    v_{2}\{8\}^{3}&\approx \overline{v}^{3}_{2} + \frac{3\overline{v}_{2}(\sigma^{2}_{y}-\sigma^{2}_{x})}{2} - (\frac{21}{11}s_{30}+3s_{12}) - \frac{-\frac{31}{11}\kappa_{40}-\frac{6}{11}\kappa_{22}+3\kappa_{04}}{4\overline{v}_{2}} \\
	  &\quad - \frac{9(\sigma^{2}_{y}-\sigma^{2}_{x})^{2}}{8\overline{v}_{2}} + \frac{\frac{20}{11}p_{50}+\frac{56}{11}p_{32}+\frac{36}{11}p_{14}}{4\overline{v}_{2}^{2}} + \frac{(\sigma^{2}_{y}-\sigma^{2}_{x})(\frac{-3}{11}s_{30}+\frac{105}{11}s_{12})}{2\overline{v}_{2}^{2}} , \quad\text{and}\\
    v_{2}\{10\}^{3}&\approx \overline{v}^{3}_{2} + \frac{3\overline{v}_{2}(\sigma^{2}_{y}-\sigma^{2}_{x})}{2} - (\frac{36}{19}s_{30}+3s_{12}) - \frac{-\frac{53}{19}\kappa_{40}-\frac{12}{19}\kappa_{22}+3\kappa_{04}}{4\overline{v}_{2}} \\
    &\quad - \frac{9(\sigma^{2}_{y}-\sigma^{2}_{x})^{2}}{8\overline{v}_{2}} + \frac{\frac{163}{95}p_{50}+\frac{94}{19}p_{32}+\frac{63}{19}p_{14}}{4\overline{v}_{2}^{2}} + \frac{(\sigma^{2}_{y}-\sigma^{2}_{x})(\frac{-6}{19}s_{30}+\frac{183}{19}s_{12})}{2\overline{v}_{2}^{2}}.
  \end{aligned}
\end{equation}
For $\vk^4$:
\begin{equation}
  \label{expand23}
  \begin{aligned}
    v_{2}\{4\}^{4}&\approx \overline{v}^{4}_{2} + 2\overline{v}^{2}_{2}(\sigma^{2}_{y}-\sigma^{2}_{x}) -4\overline{v}_{2}(s_{30}+s_{12}) - (\kappa_{40}+2\kappa_{22}+\kappa_{04}) \\
    &\quad - (\sigma^{2}_{y}-\sigma^{2}_{x})^{2} ,\\
    v_{2}\{6\}^{4}&\approx \overline{v}^{4}_{2} + 2\overline{v}^{2}_{2}(\sigma^{2}_{y}-\sigma^{2}_{x}) -4\overline{v}_{2}(\frac{2}{3}s_{30}+s_{12}) + (\kappa_{40}-\kappa_{04}) \\
    &\quad - (\sigma^{2}_{y}-\sigma^{2}_{x})^{2} + \frac{p_{50}+2p_{32}+p_{14}}{\overline{v}_{2}} - \frac{(\sigma^{2}_{y}-\sigma^{2}_{x})(\frac{4}{3}s_{30}-4s_{12})}{\overline{v}_{2}} ,\\
    v_{2}\{8\}^{4}&\approx \overline{v}^{4}_{2} + 2\overline{v}^{2}_{2}(\sigma^{2}_{y}-\sigma^{2}_{x}) -\overline{v}_{2}(\frac{28}{11}s_{30}+4s_{12}) + (\frac{31}{30}\kappa_{40}+\frac{2}{11}\kappa_{22}-\kappa_{04}) \\
	  &\quad - (\sigma^{2}_{y}-\sigma^{2}_{x})^{2} + \frac{\frac{20}{33}p_{50}+\frac{56}{33}p_{32}+\frac{12}{11}p_{14}}{\overline{v}_{2}} - \frac{(\sigma^{2}_{y}-\sigma^{2}_{x})(\frac{16}{11}s_{30}-\frac{48}{11}s_{12})}{\overline{v}_{2}} ,\quad\text{and}\\
    v_{2}\{10\}^{4}&\approx \overline{v}^{4}_{2} + 2\overline{v}^{2}_{2}(\sigma^{2}_{y}-\sigma^{2}_{x}) -\overline{v}_{2}(\frac{48}{19}s_{30}+4s_{12}) + (\frac{53}{57}\kappa_{40}+\frac{4}{19}\kappa_{22}-\kappa_{04}) \\
    &\quad - (\sigma^{2}_{y}-\sigma^{2}_{x})^{2} + \frac{\frac{163}{285}p_{50}+\frac{94}{57}p_{32}+\frac{21}{19}p_{14}}{\overline{v}_{2}} - \frac{(\sigma^{2}_{y}-\sigma^{2}_{x})(\frac{28}{19}s_{30}-\frac{84}{19}s_{12})}{\overline{v}_{2}}.
  \end{aligned}
\end{equation}
Finally, for $\vk^5$:
\begin{equation}
  \label{expand24}
  \begin{aligned}
    v_{2}\{4\}^{5}&\approx \overline{v}^{5}_{2} + \frac{5}{2}\overline{v}^{3}_{2}(\sigma^{2}_{y}-\sigma^{2}_{x}) -\overline{v}^{2}_{2}(5s_{30}+5s_{12}) - \frac{5}{4}\overline{v}_{2}(\kappa_{40}+2\kappa_{22}+\kappa_{04}) \\
    &\quad - \frac{5}{8}\overline{v}_{2}(\sigma^{2}_{y}-\sigma^{2}_{x})^{2} - (\sigma^{2}_{y}-\sigma^{2}_{x})(\frac{5}{2}s_{30}+\frac{5}{2}s_{12}) ,\\
    v_{2}\{6\}^{5}&\approx \overline{v}^{5}_{2} + \frac{5}{2}\overline{v}^{3}_{2}(\sigma^{2}_{y}-\sigma^{2}_{x}) -\overline{v}^{2}_{2}(\frac{10}{3}s_{30}+5s_{12}) + \frac{5}{4}\overline{v}_{2}(\kappa_{40}-\kappa_{04}) \\
    &\quad - \frac{5}{8}\overline{v}_{2}(\sigma^{2}_{y}-\sigma^{2}_{x})^{2} + \frac{5}{4}(p_{50}+2p_{32}+p_{14})-(\sigma^{2}_{y}-\sigma^{2}_{x})(\frac{10}{3}s_{30}-\frac{5}{2}s_{12}) ,\\
    v_{2}\{8\}^{5}&\approx \overline{v}^{5}_{2} + \frac{5}{2}\overline{v}^{3}_{2}(\sigma^{2}_{y}-\sigma^{2}_{x}) -\overline{v}^{2}_{2}(\frac{35}{11}s_{30}+5s_{12}) + \overline{v}_{2}(\frac{155}{132}\kappa_{40}+\frac{5}{22}\kappa_{22}-\frac{5}{4}\kappa_{04}) \\
	  &\quad - \frac{5}{8}\overline{v}_{2}(\sigma^{2}_{y}-\sigma^{2}_{x})^{2} + (\frac{25}{33}p_{50}+\frac{70}{33}p_{32}+\frac{15}{11}p_{14})-(\sigma^{2}_{y}-\sigma^{2}_{x})(\frac{75}{22}s_{30}-\frac{65}{22}s_{12}),\quad\text{and}\\
    v_{2}\{10\}^{5}&\approx \overline{v}^{5}_{2} + \frac{5}{2}\overline{v}^{3}_{2}(\sigma^{2}_{y}-\sigma^{2}_{x}) -\overline{v}^{2}_{2}(\frac{60}{19}s_{30}+5s_{12}) + \overline{v}_{2}(\frac{265}{228}\kappa_{40}+\frac{5}{19}\kappa_{22}-\frac{5}{4}\kappa_{04}) \\
    &\quad - \frac{5}{8}\overline{v}_{2}(\sigma^{2}_{y}-\sigma^{2}_{x})^{2} + (\frac{163}{228}p_{50}+\frac{235}{114}p_{32}+\frac{105}{76}p_{14}) - (\sigma^{2}_{y}-\sigma^{2}_{x})(\frac{65}{19}s_{30}-\frac{115}{38}s_{12}).  
  \end{aligned}
\end{equation}
  
\section{Expanding the \texorpdfstring{$h_{1}$}{h1} hydrodynamic probe to include terms up to the fifth central moment}
  \label{sec:appC}

The hydrodynamic probe, $h_{1}$, was introduced in Ref.~\cite{Giacalone:2016eyu}, but in this earlier reference its Taylor expansion was only shown to the 3rd moment of the $(v_{x},v_{y})$ distribution. If none of the terms from Eqs.~(\ref{expand1})--(\ref{expand3}) are neglected, the right-hand side of Eq.~(\ref{momHydro1}) becomes
\begin{equation}
\label{momHydro1Full}
\frac{v_{2}\{6\}-v_{2}\{8\}}{{v}_{2}\{4\}-v_{2}\{6\}} \approx \frac{1}{11}\bigg(1 - \frac{4\kappa_{40}+\frac{8(p_{50}+p_{32})}{\overline{v}_{2}}}{2\overline{v}_{2}s_{30}+3(\kappa_{40}+\kappa_{22})+\frac{3(p_{50}+2p_{32}+p_{14})-2(\sigma^{2}_{y}-\sigma^{2}_{x})(5s_{30}-6s_{12})}{2\overline{v}_{2}}}\bigg).
\end{equation}
Without neglecting any terms in Eq.~(\ref{expand21}), the full relationship corresponding to Eq.~(\ref{corrHydro1}) is
\begin{equation}
  \label{corrHydro1Full}
    \frac{v_{2}\{6\}-v_{2}\{8\}}{{v}_{2}\{4\}-v_{2}\{6\}} \approx \frac{1}{11} - \frac{1}{11}\frac{v_{2}\{4\}^{2}-12v_{2}\{6\}^{2}+11v_{2}\{8\}^{2}}{v_{2}\{4\}^{2}-v_{2}\{6\}^{2}+\frac{(\sigma^{2}_{y}-\sigma^{2}_{x})s_{30}}{3\overline{v}^{3}_{2}}}.
\end{equation}

\section{Validity of the conditions given by Eq.~(\ref{s12})}
\label{sec:appD}

Moments of the elliptic-power distribution are obtained in Ref.~\cite{Yan:2014afa}:
\begin{equation}
  \label{defMom}
  \begin{aligned}
    \sangle{ \epsilon_{n,x}^{2k}\epsilon_{n,y}^{2l} } &= X_{2k}\, \,_3\tilde{F}_2 \big(k+\frac{1}{2},\alpha+1,\alpha+\frac{1}{2};\frac{1}{2},1+l+k+\alpha;\epsilon_{0}^{2} \big),\quad\text{and} \\
    \sangle{ \epsilon_{n,x}^{2k+1}\epsilon_{n,y}^{2l} } &= \epsilon_{0} X_{2k+1}\, \,_3\tilde{F}_2 \left(k+\frac{3}{2},\alpha+1,\alpha+\frac{3}{2};\frac{3}{2},2+l+k+\alpha;\epsilon_{0}^{2} \right),
  \end{aligned}
\end{equation}
where
\begin{equation}
  \label{X2k}
\begin{aligned}
    X_{2k} &= \frac{\alpha}{\sqrt{\pi}} (1-\epsilon_{0}^{2})^{\alpha+1/2} \Gamma(\alpha) \Gamma\big(k+\frac{1}{2} \big) \Gamma\big(l+\frac{1}{2} \big),\quad\text{and} \\
    X_{2k+1} &= \frac{\alpha(1+2\alpha)}{2\sqrt{\pi}} (1-\epsilon_{0}^{2})^{\alpha+1/2} \Gamma(\alpha) \Gamma\big(k+\frac{3}{2} \big) \Gamma\big(l+\frac{1}{2} \big),
  \end{aligned}
\end{equation}
and $\,_3\tilde{F}_2$ is the regularized hypergeometric function. For the case when the ellipticity parameter tends to zero, $\epsilon_{0} \to 0$, the upper formulas give the following ratios between moments of the initial anisotropy distribution:
\begin{equation}
  \label{epsCond}
  \begin{aligned}
    \frac{\sangle{ \epsilon_{n,x}^{1}\epsilon_{n,y}^{2} }}{\sangle{ \epsilon_{n,x}^{3}\epsilon_{n,y}^{0} }} &= \frac{\Gamma \big(\frac{3}{2} \big)\Gamma \big(\frac{3}{2} \big)}{\Gamma \big(\frac{5}{2} \big)\Gamma \big(\frac{1}{2} \big)} = \frac{1}{3} , \\
    \frac{\sangle{ \epsilon_{n,x}^{2}\epsilon_{n,y}^{2} }}{\sangle{ \epsilon_{n,x}^{4}\epsilon_{n,y}^{0} }} &= \frac{\Gamma \big(\frac{3}{2} \big)\Gamma \big(\frac{3}{2} \big)}{\Gamma \big(\frac{5}{2} \big)\Gamma \big(\frac{1}{2} \big)} = \frac{1}{3},\quad\text{and} \\
    \frac{\sangle{ \epsilon_{n,x}^{1}\epsilon_{n,y}^{4} }}{\sangle{ \epsilon_{n,x}^{5}\epsilon_{n,y}^{0} }} &= \frac{\sangle{ \epsilon_{n,x}^{3}\epsilon_{n,y}^{2} }}{\sangle{ \epsilon_{n,x}^{5}\epsilon_{n,y}^{0} }} = \frac{\Gamma \big(\frac{5}{2} \big)\Gamma \big(\frac{3}{2} \big)}{\Gamma \big(\frac{7}{2} \big)\Gamma \big(\frac{1}{2} \big)} = \frac{1}{5}.
  \end{aligned}
\end{equation}
By assuming linear response between the flow and the initial anisotropy, the relations given by Eq.~(\ref{epsCond}) are also well satisfied for the corresponding moments of the $v_{n}$ distribution. Furthermore, these relations are considerably transferred to relations between the corresponding central moments of the $v_{n}$ distribution given by Eq.~(\ref{s12}). We tested these conditions numerically and found that they are satisfied to a reasonable degree in the range of the eccentricity parameter $\epsilon_{0} \le 0.15$.
 
\cleardoublepage \section{The CMS Collaboration \label{app:collab}}\begin{sloppypar}\hyphenpenalty=5000\widowpenalty=500\clubpenalty=5000
\cmsinstitute{Yerevan Physics Institute, Yerevan, Armenia}
{\tolerance=6000
A.~Tumasyan\cmsorcid{0009-0000-0684-6742}
\par}
\cmsinstitute{Institut f\"{u}r Hochenergiephysik, Vienna, Austria}
{\tolerance=6000
W.~Adam\cmsorcid{0000-0001-9099-4341}, J.W.~Andrejkovic, T.~Bergauer\cmsorcid{0000-0002-5786-0293}, S.~Chatterjee\cmsorcid{0000-0003-2660-0349}, K.~Damanakis\cmsorcid{0000-0001-5389-2872}, M.~Dragicevic\cmsorcid{0000-0003-1967-6783}, A.~Escalante~Del~Valle\cmsorcid{0000-0002-9702-6359}, P.S.~Hussain\cmsorcid{0000-0002-4825-5278}, M.~Jeitler\cmsAuthorMark{1}\cmsorcid{0000-0002-5141-9560}, N.~Krammer\cmsorcid{0000-0002-0548-0985}, L.~Lechner\cmsorcid{0000-0002-3065-1141}, D.~Liko\cmsorcid{0000-0002-3380-473X}, I.~Mikulec\cmsorcid{0000-0003-0385-2746}, P.~Paulitsch, F.M.~Pitters, J.~Schieck\cmsAuthorMark{1}\cmsorcid{0000-0002-1058-8093}, R.~Sch\"{o}fbeck\cmsorcid{0000-0002-2332-8784}, D.~Schwarz\cmsorcid{0000-0002-3821-7331}, M.~Sonawane\cmsorcid{0000-0003-0510-7010}, S.~Templ\cmsorcid{0000-0003-3137-5692}, W.~Waltenberger\cmsorcid{0000-0002-6215-7228}, C.-E.~Wulz\cmsAuthorMark{1}\cmsorcid{0000-0001-9226-5812}
% ...
\par}
\vskip\cmsinstskip
\dag:~Deceased\\
$^{1}$Also at TU Wien, Vienna, Austria\\
% ...
\end{sloppypar}
%%% END EDITABLE REGION %%%
% skeleton_end
\end{document}